\documentclass[prd,aps,preprint,tightenlines,superscriptaddress,nofootinbib,showpacs,showkeys,floatfix]{revtex4-1}

\usepackage{amsfonts}
\usepackage{pifont}    % For the to do list
\usepackage{array}
\usepackage{epsfig}
\usepackage{rotating}
\usepackage{multirow}
\usepackage{soul}
\usepackage{amsmath}    % need for subequations
\usepackage{verbatim}   % useful for program listings
\usepackage{color}      % use if color is used in text
\usepackage{colordvi}
\usepackage{hyperref}   % use for hypertext links, including those to external documents and URLs
\usepackage{tikz}
\usetikzlibrary{decorations.pathmorphing}
\usetikzlibrary{decorations.markings}
\usetikzlibrary{positioning, shapes, snakes, arrows}
\usetikzlibrary{arrows.meta}
\usepackage{caption}
\usepackage{subcaption}

\usepackage[shortlabels]{enumitem}  % needed for the todo list

\usepackage{graphicx}

\newcommand{\CC}{{\mathbb C}}
\newcommand{\QQ}{{\mathbb Q}}
\newcommand{\RR}{{\mathbb R}}
\newcommand{\sha}{\mathrel{\raisebox{\depth}{\rotatebox{270}{$\exists$}}}}
\newcommand{\ourcode}{\texttt{HyperFORM}\ }
\newcommand{\ourufcode}{\texttt{uf}\ }
\newcommand{\form}{\texttt{FORM}\ }
\newcommand{\flint}{\texttt{FLINT}\ }
\newcommand{\mincer}{\texttt{Mincer}\ }
\newcommand{\forcer}{\texttt{Forcer}\ }
\newcommand{\hyperint}{\texttt{HyperInt}\ }
\newcommand{\maple}{\texttt{MAPLE}\ }

\newcommand{\ud}{{\rm d}}
\newcommand{\ue}{{\rm e}}
\newcommand{\EulerG}{\gamma_{\rm E}}
\newcommand{\ep}{\ensuremath{\epsilon}}

\newcommand{\integrand}{\ensuremath{\mathcal{I}}}

\newcommand{\figcite}[2]{Fig.~#1 of \cite{#2}}
\newcommand{\fig}[1]{Fig.~\ref{#1}}
\newcommand{\tab}[1]{Tbl.~\ref{#1}}

\tikzset{
  quark/.style n args = {3}{postaction={decorate},
    decoration={markings,mark=at position #1 with {\arrow{latex}; \node [#3] () {#2};}}},
  quark/.default={0.5}{}{above},
  vertex/.style={circle, draw, fill=black, inner sep=1pt},
  BrownVx/.style={circle, draw, fill=red, color=red, inner sep=2pt},
  IntOutVx/.style={circle, draw, fill=gray, color=gray, inner sep=1pt},
  ChenWuLine/.style = {
    dotted
  },
  LineUnderInt/.style = {
    color=red,
    dashed,
    arrows = {Latex[width'=1pt .8, length=8pt]-Latex[width'=1pt .8, length=8pt]},
  },
  LineUnderIntLeft/.style = {
    color=red,
    dashed,
    arrows = {Latex[width'=1pt .8, length=8pt]-},
  },
  LineUnderIntRight/.style = {
    color=red,
    dashed,
    arrows = {-Latex[width'=1pt .8, length=8pt]},
  },
  IntOutLine/.style = {
    color=gray,
    dashed
  }
}

\graphicspath{{../pics/}}
\DeclareGraphicsExtensions{.eps,.ps}

\usepackage{pslatex}
\usepackage{txfonts}
\usepackage[latin1]{inputenc}
\usepackage[T1]{fontenc}

\begin{document}
\begin{titlepage}
\thispagestyle{empty}
\noindent
DESY 25-170
\hfill
November 2025 \\
\vspace{0.6cm}

\begin{center}
  {\bf \Large
  \ourcode\ -- a \form\ package for parametric integration \\[1ex] with hyperlogarithms
  }
  \vspace{1.0cm}

 {\large
   Adam Kardos$^{\, a,b}$, 
   Sven-Olaf Moch$^{\, c}$
    and
   Oliver Schnetz$^{\, c}$
   \\
 }
 \vspace{0.8cm}
 {\footnotesize{
     \begin{flushleft}       
 {\it
   $^a$ Department of Experimental Physics, Institute of Physics, Faculty of Science and Technology, 
   \\ 
   \phantom{$^a$} 
   University of Debrecen, 4010 Debrecen, PO Box 105, Hungary \\
   
   $^b$ Institute for Theoretical Physics, ELTE E\"otv\"os Lor\'and University,
   P\'azm\'any P\'eter 1/A, H--1117 Budapest, Hungary \\

   $^c$ II. Institut f\"ur Theoretische Physik, Universit\"at Hamburg,
   Luruper Chaussee 149, D-22761 Hamburg, Germany \\
 }
     \end{flushleft}
}}
  \vspace{0.8cm}
  \large {\bf Abstract}
\end{center}
  
 \noindent 
We present an implementation of algorithms for the symbolic integration of hyperlogarithms multiplied by rational functions in the computer algebra system \form. This implementation encompasses cases where hyperlogarithms have rational letters or a rational argument. It complements the previous implementation, \hyperint\!, in \maple by Erik Panzer, utilizing the advantages of \form in the efficient handling of large symbolic expressions. Among a wide range of applications, this approach enables the computation of many Feynman integrals. 
\end{titlepage}
\newpage
\setcounter{footnote}{0}
\setcounter{page}{1}
%%
%% ---------------------------------------------------------------------------
%%

%%%%%%%%%%%%%%%%%%%%%%%%%%%%%%%%%%%%%%%%%%%%%%%%%%%%%
\section{Introduction}
%%%%%%%%%%%%%%%%%%%%%%%%%%%%%%%%%%%%%%%%%%%%%%%%%%%%%
The computation of Feynman integrals at high loop orders is one of the cornerstones of perturbative quantum field theory (QFT). Among the wide variety of methods for their computation~\cite{Weinzierl:2022eaz}, those that deliver analytic results stand out, as they provide deeper insight into the mathematical structures underlying the perturbative approach. One of the key methods in this regard is the parametric integration of hyperlogarithms, which was used in the
theory of graphical functions to
obtain hallmark results for the renormalization of QFTs at the highest loop orders (the number of independent
cycles in the Feynman graph) -- namely, {\ensuremath{\phi^3}} theory at six loops~\cite{Schnetz:2025wtu} and {\ensuremath{\phi^4}} theory at seven loops~\cite{Schnetz:2022nsc}.

Parametric integration was developed by Francis Brown as a universal tool for calculating integrals \cite{Brown:2008um}.
The method is restricted to the case that after each integration step the result
can be expressed in terms of hyperlogarithms. This restriction is known under the
term linear reducability.

Brown's ideas were implemented in the computer algebra system (CAS) \maple by
Christian Bogner and by Erik Panzer in two different setups \cite{Panzer:2014caa,Bogner:2015nda}.
The implementation by Panzer -- \hyperint
\footnote{The \hyperint package for \maple\ is available under 
\url{https://bitbucket.org/PanzerErik/hyperint/wiki/Home}.} -- was particularly powerful and
suitable for calculations in QFT. It became a classical tool
for performing Feynman integrals at moderate loop order.

The drawback of parametric integration in general is that it is a rather brute-force algorithm, which generates large expressions in intermediate results (compared, e.g., with the graphical functions method, which, however, is restricted to specific Feynman graphs~\cite{Schnetz:2013hqa,Borinsky:2021gkd,Schnetz:2022nsc,Schnetz:2025opm}). 
This structural problem of parametric integration is amplified in \hyperint by its use of \maple\!, which is a good, but not the best, CAS for handling large expressions. 
Moreover, the development of \hyperint stalled soon after its first version was completed, so it was never optimized. 
Some mathematical structures that help to make parametric integration more powerful or more general did not make it into \hyperint\!. 
Still, \hyperint is a very powerful tool that is widely used by the physics community (and beyond). 

Here, we report on the first steps toward developing a more powerful parametric integration program. We chose the CAS \form~\cite{Tentyukov:2007mu,Ruijl:2017cxj} for the following three reasons. 
First, calculations with \form are fast and scale very well with large expressions. 
Second, \form is public domain\footnote{The \form program is available at \url{https://github.com/form-dev/form}.}, and third, \form scales very well on many cores, which is particularly helpful for use on modern multicore and multi-CPU servers. 
In the first step, we have translated most of the functionality of \hyperint to \form\!, which constitutes the now available first version of \ourcode\!\footnote{\ourcode\! is available at \url{https://github.com/adamkardos/HyperFORM}} \cite{kardos_2025_17706909}. We have not yet implemented further optimizations or additional functionalities in \ourcode\!; this will be part of ongoing future developments of the program. 
Since we are not introducing new mathematical ideas, we focus in this article on explaining the functionality of \ourcode and refer the reader 
for examples to Ref.~\cite{Panzer:2014gra} and for the mathematical background to Refs.~\cite{Panzer:2014caa,Brown:2008um}.

The building blocks of parametric integration are hyperlogarithms (Goncharov polylogarithms~\cite{Goncharov:1998kja}), which can be recursively defined by
\begin{equation}\label{Ldef}
    \partial_zL(a_n,a_{n-1},\ldots,a_1;z)=\frac{L(a_{n-1},\ldots,a_1;z)}{z-a_n},\quad
    \text{and}\quad L(a_n,a_{n-1},\ldots,a_1;0)=0,
\end{equation}
where $a_1,\ldots,a_n,z\in\CC$. For the evaluation of $L(a_n,a_{n-1},\ldots,a_1;z)$
we employ the regularization $\log(0)=0$,  see Section \ref{subsec:reglim}.
The definition is equivalent to
\begin{equation}
    L(\underbrace{0,\ldots,0}_n;z)=\frac{\log(z)^n}{n!}
    \label{eq:HyperlogToLogConversion}
\end{equation}
 (note that $L(0,\ldots,0;0)=0$ due to the regularization $\log(0)=0$) and
\begin{equation}
    L(a_n,a_{n-1},\ldots,a_1;z)=\int_0^z\frac{L(a_{n-1},\ldots,a_1;t)}{t-a_n}\ud t,\quad\text{if}\quad a_1\neq0,
\end{equation}
in combination with the product rule
\begin{equation}\label{shuffle}
    L(u;z)L(v;z)=L(u\sha v;z)
\end{equation}
for the sum of words $u\sha v$ which is the shuffle product of the words $u$ and $v$. 
Hyperlogarithms can also be considered as iterated integrals from $0$ to $z$ in the
sense of Chen \cite{Chen:1977oja}
\begin{equation}
    L(a_n,a_{n-1},\ldots,a_1;z)=I(z;a_n,a_{n-1},\ldots,a_1;0).
\end{equation}
Hyperlogarithms are multivalued functions. Their value depends on a choice of the path
from $0$ to $z$ which is hidden in the notation of $L(a_n,\ldots,a_1;z)$.
In the context of this article, for $z\in\RR_+$, the path is an interval on the positive real axis,
starting from the origin. This uniquely defines the sheet of $L(a_n,\ldots,a_1;z)$
and the analytic continuation away from the positive real axis.

If the numbers (letters) $a_1,\ldots,a_n$ are $0$ or $1$, then the hyperlogarithm
is a multiple polylogarithm. Multiple polylogarithms evaluate at $1$ (and at $\infty$)
to multiple zeta values (MZV), which are $\QQ$ linear combinations of multiple zeta sums
\begin{equation}\label{mzv}
    \zeta(n_r,\ldots,n_1)=\sum_{1\leq k_1<\ldots<k_r}\frac1{k_1^{n_1}\cdots k_r^{n_r}}
    ,\quad r\geq1,\;n_r\geq2.
\end{equation}
Multiple zeta values fulfill many relations (such as $\zeta(2,1)=\zeta(3)$).
Conjecturally, all relations between MZVs are generated by (quasi-)shuffle relations \cite{Blumlein:2009cf}. For high weight $n_1+\ldots+n_r$, it is much more convenient to utilize
the motivic structure of MZVs for their reduction to a $\QQ$ basis \cite{Brown:2011uzk}.
Here, we only need low weights where it is convenient to collect the relations between
MZVs in a table. Note that the motivic reduction of MZVs is implemented in \form.

%%%%%%%%%%%%%%%%%%%%%%%%%%%%%%%%%%%%%%%%%%%%%%%%%%%%%
\section{Implementation}
\label{sec:impl}
%%%%%%%%%%%%%%%%%%%%%%%%%%%%%%%%%%%%%%%%%%%%%%%%%%%%%
We chose to implement hyperlogarithms in the CAS \form~\cite{Tentyukov:2007mu,Ruijl:2017cxj}, which is open source, capable of handling very large expressions, and available for Unix-based operating systems. 
Furthermore, being open source, it can be compiled for all major architectures (x64 and ARM), and it offers extended support for multithreaded execution.
\form\ also supports polynomial operations like
the greatest common divisor (GCD) both natively
or through an interface to \flint\ \cite{flint, 7891956}. 
This is beneficial because
manipulations of polynomials and rational functions are necessary at several steps
during parametric integration.

In QFT, Feynman integrals often need regularization, with dimensional regularization in $D=2n-2\ep$ dimensions and $n\in \mathbb{N}$ being the method of choice (usually in $D=4-2\ep$). 
\ourcode\ is designed for hyperlogarithm-based integration of Laurent expansions in $\ep$ of integrals of the type
\begin{align}
\label{eq:integral-def}
    I(\ep) &=
    \int_{0}^{\infty}\ud\, x_1\cdots\int_{0}^{\infty}\ud\, x_N
    \prod_{j=1}^{M} f_j^{n_j + m_j \ep}(x_1,\dots, x_N)
    \,,\quad
    n_j\in\mathbb{Z},\,m_j\in\mathbb{Q}
    \,,
\end{align}
where the $f_j$ are rational functions of the integration variables.

To use \ourcode, the user should create a text file which starts with the header
\begin{verbatim}
    #include- hyperform.h
\end{verbatim}
The package consists of three files:
\begin{itemize}
    \item \texttt{hyperform.h}: contains all procedures of the package.
    \item \texttt{declare-hyperform.h}: contains all symbol, function and
    preprocessor variable declarations. The user should not change the content of this file.
    Changing a preprocessor variable (if necessary) is better done in the user source file with a \texttt{\#redefine} statement \emph{after} including the
    \ourcode\ source.
    \item \texttt{mzvlow.h}: contains a tabulated reduction of MZVs taken from \form's 
    {\it Multiple Zeta Values Data Mine} \footnote{See \url{https://www.nikhef.nl/~form/datamine/datamine.html}. \label{footnote-mzvlow}},  which is described in Ref.~\cite{Blumlein:2009cf}. 
\end{itemize}
The input to \ourcode consists of the integrands in Eq.~(\ref{eq:integral-def}). The user is free to choose the names for $\ep$ and for the integration variables. 
The only restriction on the user-defined name convention and the form of the input is dictated by
the allowed syntax of the language of \form. 
An integrand \texttt{`IntegralExpr'} takes the following form
\begin{align}
\label{eq:integrand-example}
\texttt{`IntegralExpr'}
    &= \frac{
    (f_1)^{k_1}\cdots (f_M)^{k_M}
    }{
    (g_1)^{l_1}\cdots (g_N)^{l_N}
    }
    (h_1)^{n_1 + m_1\ep}\cdots (h_L)^{n_L + m_M\ep}
    \,,\quad
    k_j,l_j \in \mathbb{N}, n_j\in\mathbb{Z},\,m_j\in\mathbb{Q} \backslash \{0\}
    \,.
\end{align}
In \ourcode\!, the polynomials $f_j$, $g_j$ and $h_j$ have the (preferred) input syntax
\begin{verbatim}
   'IntegralExpr' = num(f1)^k1*...*num(fM)^kN*
      den(...)^l1*...*den(...)^lN*
      (...)^(n1 + m1*ep)*...*(...)^(nM + mM*ep);
\end{verbatim}
where the brackets contain the polynomials $h_j$ and the functions \texttt{num} and \texttt{den} 
accept those polynomials $f_j$ and $g_j$, respectively, as arguments, which appear in Eq.~(\ref{eq:integrand-example}) without regularizing \ep-powers in the exponent.\footnote{For a detailed discussion of the syntax for rational polynomials in \form\!, in particular for expressions of the type $1/(g_j)^{k_j}$ with integer $k_j >0$, see the \form manual, Sec.~7.32 on denominators.
\url{https://github.com/vermaseren/form/releases/download/v4.3.1/form-4.3.1-manual.pdf}
}
After the definition of the integrand, a call to \texttt{HypParseInputExpr} will translate
the user-defined variables and functions to \ourcode\!'s internal notation. 
Assuming, for example, that the user has defined the integration variables $x,y,z,u,v,w$, 
the \ep-symbol \texttt{ep}, the numerators and the denominators in the integrand as functions \texttt{num} and \texttt{den}, respectively, 
then the call to the input parser routine takes the form:
\begin{verbatim}
    #call HypParseInputExpr(ep,num,den,x,y,z,u,v,w)
\end{verbatim}
In order to avoid collision between the user's and \ourcode\!'s internal name-space,
all internal symbols, functions and procedures have a \texttt{HYP} or \texttt{Hyp} prefix, respectively.
This gives more freedom for importing integrands.

The integral can be finite or singular in the limit $\ep\to0$.
In the first case, we call the integral \ep-finite; this is the simplest situation.
In the latter case, a regularization procedure
is necessary prior to integration. In \ourcode\ we adopted the regularization procedure of
Ref.~\cite{Panzer:2014gra}. 
In a future version of \ourcode\ we will implement a more efficient regularization procedure.

The \ep\ regularization is done by a call
\begin{verbatim}
    #call HypAutoRegularize('IntegralExpr',...)
\end{verbatim}
where the ellipses represent additional variables in the integrand that are not integrated over.
The regularization procedure is recursive and automatic; it finishes when no further 
divergence is found. 
Details are given in Sec.~\ref{sec:regularize} and an example is presented in eqs.~\eqref{eq:twomassTri} and \eqref{eq:twomassTri-reg}.
The algorithm also attempts to find the shortest regularization.

After regularization, we can expand the integrand in \ep. This is done by a call
\begin{verbatim}
    #call HypEpExpand
\end{verbatim}
The procedure \texttt{HypEpExpand} has no argument. The maximum power of \ep\ 
is determined by the preprocessor variable \texttt{HYPMAXEP} (by default set to zero).

In case of projective integrals (like parametric Feynman integrals)
we need to specify the affine sheet by setting one integration variable to $1$.
In physics literature~\cite{Weinzierl:2022eaz}, this is known as the Chen-Wu theorem.
A call for this is for example
\begin{verbatim}
    #call HypApplyChenWu(F1,3)
\end{verbatim}
where we set the third integration variable to $1$ in the integrand given by the local
expression \texttt{F1}. 
It is advisable to use the \texttt{HypApplyChenWu} command for this purpose to avoid error messages from automated checks regarding the uniformity of the integrand.

In the current version of the code, it is possible to detect coefficients
of hyperlogarithms that non-trivially equal $0$. This eliminates terms in intermediate steps that cancel in the final result. This option is activated by default; it can be turned off by
\begin{verbatim}
    #redefine HYPcheckZeroCoefficients "0"
\end{verbatim} 
after including the source for \ourcode\!.

In the next step, it is recommended to define an integration sequence.
The choice of an integration sequence is important in two ways.
Firstly, more complex integrals may have some integration sequences that are linearly reducible
(and hence tractable with \ourcode) while others are not.
Secondly, different linearly reducible integration sequences can have very different
runtime. So, for complex problems, the user is advised to test several integration sequences
before spending resources on one specific integration sequence. 
For integration variables \texttt{y1} \dots, \texttt{y5}, the call is, for example,
\begin{verbatim}
    #define IntegrationSequence "3,4,5,2,1"
\end{verbatim} 
if one wants to integrate with respect to \texttt{y3} first, followed by
\texttt{y4}, \texttt{y5}, \texttt{y2}, and \texttt{y1}.
The actual integration over the $i$th integration variable for an integrand
in the local expression \texttt{F1} is performed by a call to the \texttt{HypIntegrationStep} 
routine:
\begin{verbatim}
      #call HypIntegrationStep(F1,`i')
\end{verbatim}
If the user has defined the variable \texttt{IntegrationSequence},
integration is a do-loop through the specified sequence
\begin{verbatim}
    #do IntVar={`IntegrationSequence'}
      #call HypIntegrationStep(F1,`IntVar')
    #enddo
\end{verbatim}
To get the most aesthetic result, it is recommended that the user ends the calculation
with a call of \texttt{HypFinalizeResult}
\begin{verbatim}
    #call HypFinalizeResult(ep,`HYPMAXEP')
\end{verbatim}
In this call, the user specifies the \ep\ symbol they want to use in the final printout 
and the order of the expansion in \ep.
This is necessary because it can happen that a parametric integral
gets multiplied by an \ep-singular $\Gamma$ function. 
In such a case, the final result has lower order in \ep\ than specified at the beginning of the calculation.

%----------------------------------------
\subsection{Installation}
%----------------------------------------
The code is available from its \texttt{github} repository:
\begin{verbatim}
    git clone https://github.com/adamkardos/HyperFORM.git
\end{verbatim}
It needs a recent version of \form\footnote{We have tested 
\ourcode with the \form\ version available in the commit of \texttt{2073865}.}.
Running examples and the test suite requires a \texttt{FORMPATH}
set up to include the installation folder of the package:
\begin{verbatim}
    # in your ~/.bashrc or ~/.zshrc file:
    FORMPATH="/path/to/HyperFORM/src:$FORMPATH"
    export FORMPATH
\end{verbatim}
The package comes with an extensive set of tests. The user is advised to
run the tests prior to using the code in order to check compatibility with the installed version of \form\footnote{\ourcode 
assumes the use of \flint\ \cite{flint, 7891956} for polynomial operations. If this is not
the case, checks with GCD and some other operations will fail due to different 
normalization conventions between \form\ and \flint.}. 

Running the tests is easy due to the standard \texttt{Ruby} script created for \form:
\begin{verbatim}
    cd check
    ./check.rb
\end{verbatim}
If necessary, the user can check the functionality of a single routine by:
\begin{verbatim}
    cd check
    ./check.rb HypRegularizeAndDifferentiateForFibrationBases.frm
\end{verbatim}
%

%%%%%%%%%%%%%%%%%%%%%%%%%%%%%%%%%%%%%%%%%%%%%%%%%%%%%
\section{Examples}
%%%%%%%%%%%%%%%%%%%%%%%%%%%%%%%%%%%%%%%%%%%%%%%%%%%%%
In the following, we illustrate the use of \ourcode\ by showcasing
its application to various problems related to perturbative QFT.

%----------------------------------------
\subsection{Regularized limits}
\label{subsec:reglim}
%----------------------------------------
Handling limits is a central mathematical operation in parametric integration.
The original expression for the integral is finite, so handling of singularities does
not seem necessary. However, parametric integration splits the full expression into a sum of terms
that (by linearity of the algorithm) are treated separately. Each individual term can be singular;
only the sum of all terms is guaranteed to be finite. The type of singularities one encounters is
$\lim_{x\to0}\log(x)^k$ and $\lim_{x\to\infty}\log(x)^k$ for positive integer $k$.
Mathematically, such a singularity arises when the boundary of an integral
intersects the singularity of the integrand. The simplest examples are
\begin{equation}\label{regex1}
\int_0^2\frac{\ud x}x\equiv\lim_{\epsilon\to0}\int_\epsilon^2\frac{\ud x}x=\log2-\lim_{\epsilon\to0}\log\epsilon,\quad \int_2^\infty\frac{\ud x}x\equiv\lim_{\delta\to\infty}\int_2^\delta\frac{\ud x}x=\lim_{\delta\to\infty}\log(\delta)-\log2,
\end{equation}
where the limits do not exist.
It is possible to keep these limits unevaluated and check that they drop out in the final result.

Because the situation arises very frequently in the theory of hyperlogarithms and
MZVs (\ref{mzv}), Pierre Deligne came up with a more elegant solution \cite{Deligne1989}
that is now commonly used in the field. 
The insight is, that variable transformations (in particular $x\mapsto\lambda x$) of individual 
terms are typically not used to manipulate these integrals.
This enables one to formally write $\lim_{\epsilon\to0}\log\epsilon$ as $\log0$ and
$\lim_{\delta\to\infty}\log\epsilon$ as $\log\infty$ yielding
$$
\int_0^2\frac{\ud x}x\equiv\log2-\log0,\quad \int_2^\infty\frac{\ud x}x\equiv\log(\infty)-\log2.
$$
Note that $\log 0$ and $\log\infty$ should not be read as the logarithm function applied
to $0$ and $\infty$; these evaluations do not exist. They are mere names for constants that
are treated as variables.

Because $2\cdot 0=0$, a scale transformation $x\mapsto 2x$ is clearly not consistent with the
notation:
$$
\int_0^2\frac{\ud x}x\neq\int_0^1\frac{\ud x}x,\quad \int_2^\infty\frac{\ud x}x\neq\int_1^\infty\frac{\ud x}x.
$$
Such a transformation demands one to use the $\epsilon$, $\delta$ representation in (\ref{regex1}),
where the variable change can be traced back from $\epsilon\mapsto\epsilon/2$, $\delta\mapsto\delta/2$.

The situation is even more subtle. One has to specify the direction in which the path of
integration from 0 to 2 (or from 2 to $\infty$) approaches the singular points.
In the examples, one could tacitly assume that the integration is on the positive real axis,
but even the tiniest detour to the negative numbers when approaching the singularities
(approaching $-0$ and $-\infty$) gives a $\log(-1)=\pm i\pi$ (depending on the exact shape of the
integration contour). We need a notion of homology with a `tangential base point' \cite{Deligne1989}.
The typical convention is that $0$ is approached from the
right and $\infty$ is approached from the left ($\epsilon>0$, $\delta>0$ in (\ref{regex1})).

The result of a calculation with this limited set of integration rules provides an expression
that is a polynomial in $\log(0)$ and $\log(\infty)$.
If the original integral was finite, then the result is constant in (independent of) $\log(0)$ and $\log(\infty)$.
We e.g. have
$$
\int_1^2\frac{\ud x}x=\int_0^2\frac{\ud x}x-\int_0^1\frac{\ud x}x=\log2-\log0-(\log1-\log0)=\log2.
$$
By default, this consistency condition is checked in {\tt HyperFORM}.

Note that, a priori, $\log(0)$ and $\log(\infty)$ are independent. However, it is
possible to include the inversion $x\mapsto1/x$ to the set of allowed integration rules which
links $\log(0)$ to $\log(\infty)=\log(1/0)=-\log(0)$ ($\delta=1/\epsilon$ in (\ref{regex1})).
With this extension, we e.g.\ get
$$
\int_2^\infty\frac{\ud x}x=-\int_{1/2}^0\frac{\ud x}x=\int_0^{1/2}\frac{\ud x}x=\log\frac12-\log0=-\log2-\log0.
$$
Because the initial integral is finite, one can go one step further.
One can set $\log0$ (and $\log\infty$ if one does not use the inversion) to any number in $\CC$.
Typically, one uses $\log0=\log\infty=0$ because this choice gives the simplest expressions.
With this `regularization' we write
$$
\int_0^2\frac{\ud x}x=\log2,\quad \int_2^\infty\frac{\ud x}x=-\log2.
$$
This regularization is used to extend MZVs to singular integrals. With (\ref{Ldef}), we obtain for
the convergent integral $L(0,0,1;1)$ the value $-\zeta(3)$. The regularization is compatible
with the shuffle product (\ref{shuffle}). From $L(0;1)=L(0,0;1)=0$ we obtain
\begin{align*}
&0=L(0;1)L(0,1;1)=2L(0,0,1;1)+L(0,1,0;1),\\
&0=L(0,0;1)L(1;1)=L(1,0,0;1)+L(0,1,0;1)+L(0,0,1;1).
\end{align*}
This leads to
$$
L(0,1,0;1)=2\zeta(3),\quad L(1,0,0;1)=-\zeta(3),
$$
extending the evaluations at $1$ of multiple polylogarithms to all words in 0 and 1.

Finally, it is important to notice that one can use the full set of integration rules
for divergent terms in a convergent integral if one uses them consistently.
This can be convenient in some setups to map evaluations at $\infty$ to evaluations at $0$ with
other transformations than the inversion $x\mapsto1/x$. We e.g.\ have
$$
\int_1^2\frac{\ud x}x=\int_1^\infty\frac{\ud x}x-\int_2^\infty\frac{\ud x}x=-\log(1)-(-\log2)=\log2.
$$
A transformation $x\mapsto2/x$ gives
$$
\int_1^\infty\frac{\ud x}x-\int_2^\infty\frac{\ud x}x=\int_0^2\frac{\ud x}x-\int_0^1\frac{\ud x}x=\log2.
$$
Note that one gets the false result 0 if one (inconsistently) uses the transformation $x\mapsto1/x$ for
the first term and $x\mapsto2/x$ for the second term.

In this article, we understand limits as regularized limits in the sense of this subsection.
%----------------------------------------
\subsection{Conversion to a fibration basis}
\label{subsec:FibrationBasis}
%----------------------------------------
The integration domain is the positive real axis. This means that hyperlogarithms are evaluated at $0$ and at $\infty$. While the limit at $0$ gives trivial results
by Eq.~(\ref{Ldef}), the limit at $\infty$ is non-trivial. It renders the program
with expressions that have to be converted before the next integration step.
By Eq.~(\ref{Ldef}), integrating a hyperlogarithm with respect to $z$ demands that the
integrand is given as $L(a_n,\ldots,a_1,z)$ where the constants $a_i$ do not depend on $z$.
If the limit at $\infty$ provides a hyperlogarithm with constants that are rational
functions in $z$, then this conversion is always possible.
We refer to this operation as conversion to the fibration basis. \footnote{Strictly speaking,
the term fibration basis refers to all the remaining variables not just to the next one in the integration sequence.}

Conversion of the hyperlogarithms to the fibration basis can be done only for the next integration 
variable or for all the remaining integration variables. In the current version of the code, when performing an integration sequence over an ordered
set of integration variables, the fibration basis operation is only partially carried out:
hyperlogarithms whose letters depend on the next integration variable are transformed so that this dependence is removed from the letters.
In the code there is a stand-alone
routine which can be used to perform a partial or full fibration basis conversion.

To illustrate the conversion to the fibration basis, consider the example 
presented in Eq.~(D.56) of Ref.~\cite{Anastasiou:2013srw} (where we correct a typo
by reversing the order of the hyperlogarihm):
\begin{align}
  F=
  {\rm reg\, lim}_{x_1\to \infty}
  L\left(
    -\frac{1 + x_2 + x_3}{x_3},-1
    ;
    x_1
  \right)
  \,,
\end{align}
in this expression ${\rm reg\, lim}_{x_1\to \infty}$ stands for the regularization discussed in
the previous (A) section. 
If the integration order is $x_2,\, x_3$, the fibration basis
can be calculated with the following short \form\ code using the 
procedure \texttt{HypFibrationBasis}:
\begin{verbatim}
    #include- hyperform.h
    *
    cfunctions Linf, L, rat;
    symbols x2, x3;
    *
    local F = Linf(rat(-1-x2-x3, x3),-1);
    .sort
    #call HypFibrationBasis(F,Linf,L,rat,x2,x3)
    *
    print +s;
    .end
\end{verbatim}
The procedure \texttt{HypFibrationBasis} takes as arguments (in that order) the user-defined names of the local expression \texttt{F}, the hyperlogarithms at $\infty$ \texttt{Linf}, the hyperlogarithms $L$ 
and the rational function collecting variables in the argument of the hyperlogarithms, \texttt{rat}.
This is followed by the list of the integration variables \texttt{x2}, \texttt{x3}, starting with the next integration variable, here \texttt{x2}. 
A detailed discussion of the procedure \texttt{HypFibrationBasis} and its arguments is given in the appendix.
With the \texttt{print +s;} command, \ourcode\ gives for the above example the following result:
\begin{verbatim}
   F =
       - z2
       - L(-1, - 1 - x3,x2)
       - L(-1,x2)*L(-1,x3)
       + L(-1,x2)*L(0,x3)
       + L(0,-1,x3)
       - L(0,0,x3)
      ;
\end{verbatim}
where $\mathtt{z2} = \zeta(2)$.
The result agrees with Eq.~(D.56) of \cite{Anastasiou:2013srw}.
Notice that the first term in Eq.~(D.56) of \cite{Anastasiou:2013srw}, 
$L(0,0;x_1)$, is absent in our result because its regularized limit is zero ($L(0,0;\infty)=\log(\infty)^2/2!$ and $\log(\infty)=\log(1/0)=-\log(0)=0$ by definition).

%----------------------------------------
\subsection{The calculation of $p$-integrals}\label{pint}
%----------------------------------------
\label{sec:FAtopo}
\begin{figure}[ht]
    \centering
    \begin{tikzpicture}
      \draw[quark] (0,0) node[] (in) {} --++ (1.0, 0) node[] (vx left) {};
      \draw[quark={0.5}{$q_1$}{below left}] (vx left.center) arc[start angle=180, end angle=270, x radius=12mm, y radius=12mm] --++ (0.5,0) node[] (vx bottom) {};
      \draw[quark={0.5}{$q_2$}{below right}] (vx bottom.center) --++ (0.5,0) arc[start angle=270, end angle=360, x radius=12mm, y radius=12mm] --++ (0,0) node[] (vx right) {};
      \draw[quark={0.5}{$q_3$}{above right}] (vx right.center) arc[start angle=0, end angle=90, x radius=12mm, y radius=12mm] --++ (0,0) node[] (vx top right) {};
      \draw[quark={0.5}{$q_4$}{above}] (vx top right.center) --++ (-1.0,0) node[] (vx top left) {};
      \draw[quark={0.5}{$q_5$}{above left}] (vx top left.center) arc[start angle=90, end angle=180, x radius=12mm, y radius=12mm] --++ (0,0);
      \draw[quark={0.5}{$q_7$}{right}] (vx top right.center) -- (vx bottom.center);
      \draw[quark={0.5}{$q_6$}{left}] (vx top left.center) -- (vx bottom.center);
      \draw[quark] (vx right.center) --++ (1.0, 0) node[] (out) {};
      \node[left] at (in) {$p$};
      \node[right] at (out) {$p$};
    \end{tikzpicture}
    \caption{The two-point function of the \texttt{FA} topology with external momentum $p^2 \neq 0$ and labels for all propagators.}
    \label{fig:FA-topology}
\end{figure}
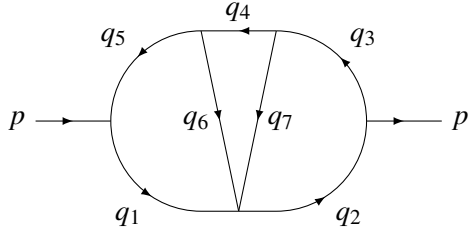

As an example of propagator ($p$)-type integrals, we calculate the three-loop Feynman diagram depicted in Fig.~\ref{fig:FA-topology}.
The graph is commonly referred to as the FA topology~\cite{Larin:1991fz},
given by the following integral in $D=4-2\epsilon$ dimension,
\begin{equation}
    \texttt{FA} = \int d^D\ell_{1}\, d^D\ell_{2}\, d^D \ell_{3}\, \frac{1}{q_1^2 \dots q_7^2}\,.
\end{equation}
In terms of the external momentum $p$ and the loop momenta $\ell_{j}$, our propagators are defined as
\begin{align}
    q_1 &= \ell_{1} + p
    \,,\quad
    q_2 = \ell_{12} + p
    \,,\quad
    q_3 = \ell_{12}
    \,,\quad
    q_4 = \ell_{13}
    \,,\quad
    q_5 = \ell_{1}
    \,,\quad
    q_6 = \ell_{3}
    \,,\quad
    q_7 = \ell_{2} - \ell_{3}
    \,,
\end{align}
with the short-hand $\ell_{ij} = \ell_{i} + \ell_{j}$. 
In terms of the associated two Symanzik polynomials, the integral reads
\begin{align}
\label{eq:FAgraphpoly}
    \texttt{FA} &= \left(p^2\right)^{-2-6\epsilon}
    \int_{0}^{\infty}\ud x_{1}\cdots \int_{0}^{\infty}\ud x_{7}\, 
  \bigl(
      x_5 x_6 x_7 + x_4 x_6 x_7 + x_3 x_6 x_7 + x_2 x_6 x_7 + x_1 x_6 x_7 + x_4 x_5 x_7
    + x_3 x_5 x_7 
\nonumber \\ &  \quad    
    + x_2 x_5 x_7 + x_1 x_4 x_7 + x_1 x_3 x_7 + x_1 x_2 x_7 + x_3 x_5 x_6 
    + x_2 x_5 x_6 + x_3 x_4 x_6 + x_2 x_4 x_6 + x_1 x_3 x_6 + x_1 x_2 x_6 
\nonumber \\ &  \quad    
    + x_3 x_4 x_5 
    + x_2 x_4 x_5 + x_1 x_3 x_4 + x_1 x_2 x_4
  \bigr)^{-1+4 \epsilon} 
  \bigl(
      x_2 x_5 x_6 x_7 + x_1 x_5 x_6 x_7 + x_2 x_4 x_6 x_7 + x_1 x_4 x_6 x_7 
\nonumber \\ &  \quad       
      + x_2 x_3 x_6 x_7      
    + x_1 x_3 x_6 x_7 + x_2 x_4 x_5 x_7 + x_1 x_4 x_5 x_7 + x_2 x_3 x_5 x_7 + x_1 x_3 x_5 x_7
    + x_1 x_2 x_5 x_7 + x_1 x_2 x_4 x_7 
\nonumber \\ &  \quad     
    + x_1 x_2 x_3 x_7 + x_2 x_3 x_5 x_6 + x_1 x_3 x_5 x_6
    + x_1 x_2 x_5 x_6 + x_2 x_3 x_4 x_6 + x_1 x_3 x_4 x_6 + x_1 x_2 x_4 x_6 + x_1 x_2 x_3 x_6
\nonumber \\ &  \quad     
    + x_2 x_3 x_4 x_5 + x_1 x_3 x_4 x_5 + x_1 x_2 x_4 x_5 + x_1 x_2 x_3 x_4
  \bigr)^{-1-3 \epsilon}\, \delta(x_i-1)\, .
 \end{align}   
For the computation of the Symanzik polynomials we have used the \form\ package \ourufcode \cite{adamkardos_2025_17646663}.

Let us discuss the corresponding \form\ program step by step.
The header of the calculation with \ourcode\ is
\begin{verbatim}
#include- hyperform.h

#redefine HYPMAx_EP "2"

#define IntegralExpr "FA"
#define IntegrationSequence "2,3,7,4,6,5"
#define ChenWuVar "1"
\end{verbatim}
where we have defined a preprocessor variable \texttt{IntegralExpr} for the name of the integrand \texttt{FA}, 
the integration sequence through a list of ordinal numbers, as discussed above, 
and a preprocessor variable \texttt{ChenWuVar}, which will be substituted later into the procedure \texttt{HypApplyChenWu} to set the first integration varable to 1, since we apply \ourcode\ to a projective Feynman integral in the parametric space.
We have also redefined the built-in preprocessor variable \texttt{HYPMAXEP} to 2, 
so that the output for \texttt{FA} at the end will be accurate including terms of $\mathcal{O}(\ep^2)$
in the parameter for dimensional regularization.

The next part consists of all declarations and the input for the integrand \texttt{IntegralExpr} in terms of the Symanzik polynomials for the Feynman diagram, see, for example, Ref.~\cite{Gluza:2010rn, Weinzierl:2022eaz}. 
\begin{verbatim}
symbols x1,...,x7;
symbol ep;
cfunction num, den;
cfunctions Gamma, invGamma;

local `IntegralExpr' = 
  (
      x5*x6*x7 + x4*x6*x7 + x3*x6*x7 + x2*x6*x7 + x1*x6*x7 + x4*x5*x7
    + x3*x5*x7 + x2*x5*x7 + x1*x4*x7 + x1*x3*x7 + x1*x2*x7 + x3*x5*x6 
    + x2*x5*x6 + x3*x4*x6 + x2*x4*x6 + x1*x3*x6 + x1*x2*x6 + x3*x4*x5 
    + x2*x4*x5 + x1*x3*x4 + x1*x2*x4
  )^(-1+4*ep)*
  (
      x2*x5*x6*x7 + x1*x5*x6*x7 + x2*x4*x6*x7 + x1*x4*x6*x7 + x2*x3*x6*x7
    + x1*x3*x6*x7 + x2*x4*x5*x7 + x1*x4*x5*x7 + x2*x3*x5*x7 + x1*x3*x5*x7
    + x1*x2*x5*x7 + x1*x2*x4*x7 + x1*x2*x3*x7 + x2*x3*x5*x6 + x1*x3*x5*x6
    + x1*x2*x5*x6 + x2*x3*x4*x6 + x1*x3*x4*x6 + x1*x2*x4*x6 + x1*x2*x3*x6
    + x2*x3*x4*x5 + x1*x3*x4*x5 + x1*x2*x4*x5 + x1*x2*x3*x4
  )^(-1-3*ep)
;
.sort

#call HypParseInputExpr(ep,num,den,x1,...,x7)
.sort
\end{verbatim}
The integration variables are \texttt{x1,\dots,x7}. 
The roles of \texttt{ep}, \texttt{num} and \texttt{den} has been discussed above and the two functions \texttt{Gamma} and \texttt{invGamma} will be used for the $\Gamma$-function and its inverse.
The two brackets in the local expression \texttt{IntegralExpr} contain the two Symanzik polynomials given in Eq.~\eqref{eq:FAgraphpoly}.
Then, the integrand is turned into
\ourcode\!'s internal representation with a call to \texttt{HypParseInputExpr}, as explained above.

The next part of the code performs an $\ep$ regularization of the integrand using the method outlined in Ref.~\cite{Panzer:2014gra}.
\begin{verbatim}
#call HypAutoRegularize(`IntegralExpr')

#call HypEpExpand

* Applying the Chen-Wu theorem on `ChenWuVar':
#call HypApplyChenWu(`IntegralExpr',ChenWuVar)
.sort
\end{verbatim}
This is achieved by calling the procedure \texttt{HypAutoRegularize}.
The integrand does not change upon calling \texttt{HypAutoRegularize}, since the \texttt{FA} integral in Eq.~\eqref{eq:FAgraphpoly} is finite, and this is confirmed by the procedure. 
Then,
the $\ep$ regularized integrand is expanded in $\ep$ with a call to \texttt{HypEpExpand}. 
Afterwards, the call to \texttt{HypApplyChenWu} substitutes $1$ for \texttt{x1} according to the choice for the preprocessor variable \texttt{ChenWuVar} above.

The actual integrations are then performed by a series of calls to the \texttt{HypIntegrationStep} procedure where
we specify the integrand expression and the actual integration variable, here \texttt{IntegralExpr} and \texttt{IntVar}:
\begin{verbatim}
#do IntVar={`IntegrationSequence'}
  #call HypIntegrationStep(`IntegralExpr',`IntVar')
#enddo
\end{verbatim}
The integration is performed in a loop over the preprocessor variable \texttt{IntVar}\footnote{
More precisely, the integration is performed for \texttt{HYPz`IntVar'}, which is the internal variable associated with \texttt{`IntVar'}. If the command \texttt{print;} is used at intermediate stages, the displayed expressions will therefore contain terms involving \texttt{HYPz`IntVar'}.}, which assumes the values defined in \texttt{IntegrationSequence} 
defined above.

The final part of the code prepares the output of the result.
\begin{verbatim}
************************
* FINAL NORMALIZATION: *
************************
multiply Gamma(1 + 3*ep);

#call HypToGscheme(3,Gamma,invGamma,ep,`HYPMAXEP')

#call HypFinalizeResult(ep,`HYPMAXEP')

#call HypPrintStatistics
.sort

bracket ep;
print +s;
.end
\end{verbatim}
Here, the command \texttt{multiply Gamma(1 + 3*ep);} multiplies the result with the $\Gamma$ function that originates from the parametric representation.
The procedure \texttt{HypToGscheme} prepares the output in the chosen scheme. 
Here, we give the result in the G-scheme, as is common practice, e.g., in the \mincer\ \cite{Larin:1991fz} and \forcer\ \cite{Ruijl:2017cxj} programs. 
A detailed explanation of the arguments for the call to the procedure \texttt{HypToGscheme} is given in the appendix.
Finally, the call to \texttt{HypFinalizeResult} performs the series expansions in $\ep$, including the $\Gamma$ functions. 
Since the $\Gamma$-function in this case does not contain any poles, the result is valid to the order in $\ep$ specified by the preprocessor variable \texttt{HYPMAXEP}.
Had the $\Gamma$ function been divergent in $\ep$, the result would only be valid up to $\mathcal{O}(\ep^{\mathtt{HYPMAXEP - 1}})$.

The call to \texttt{HypPrintStatistics} prints out some (limited) statistics of the calculation. 
When all instructions are executed successfully, the code provides the result for the integral up to $\mathcal{O} (\ep^2)$ terms with the commands \texttt{bracket ep;} for bracketing in $\ep$ and \texttt{print +s;} as
\begin{verbatim}
    FA =
       + ep * (
          + 50*z6
          - 80*z5
          - 4*z3^2
          )
       + ep^2 * (
          - 200*z6
          + 359*z7
          + 80*z5
          - 12*z3*z4
          + 16*z3^2
          )
       + 20*z5
         ;
\end{verbatim}
In this print out $\texttt{ep}=\ep$ and $\texttt{zi}=\zeta(i)$.
The result,
written out for clarity,
\begin{equation}
    \texttt{FA} = \left(p^2\right)^{-2-6\epsilon} \left(
    20 \zeta_5 + \epsilon\left(
          - 80 \zeta_5
         - 4 \zeta_3^2     
          + 50 \zeta_6
    \right) + \epsilon^2\left(
            80 \zeta_5
          + 16 \zeta_3^2     
          - 200 \zeta_6
          - 12 \zeta_3\zeta_4
          - 359 \zeta_7
    \right)
    \right)
    \, ,
\end{equation}
agrees, of course, with \mincer~\cite{Larin:1991fz}, \forcer~\cite{Ruijl:2017cxj}, \hyperint~\cite{Panzer:2014caa} and  {\texttt{HyperlogProcedures}}~\footnote{See \url{https://github.com/oliverschnetz/HyperlogProcedures}.
%csm \url{https://faubox.rrze.uni-erlangen.de/dl/fi5WHW6DDa71wcEDjUYPhu/HyperlogProcedures08.zip}
.}.

The corresponding timings can 
be found in \tab{tbl:FAtimings}. The comparison of the timings is tricky.
While \forcer and \mincer basically look up the result in a table, 
\hyperint, \ourcode, and {\tt HyperlogProcedures} perform the actual calculation.
Still, a comparison with the latter is not quite fair in some sense, because
{\tt HyperlogProcedures} uses the theory of graphical functions which fundamentally differs
from parametric integration \cite{Schnetz:2013hqa,Brown:2015ztw,Schnetz:2016fhy,Borinsky:2021jdb,Borinsky:2021gkd,Borinsky:2022lds,Schnetz:2022nsc,Schnetz:2025wtu,Schnetz:2025opm}.
In particular, the theory of graphical functions completely avoids
working with many variables or parameters and hence is much faster, much less memory consuming,
and scales much better with the loop order. The drawback of the theory of graphical functions
is that it can only be used in very special setups (massless theories with $\leq3$ external legs or vertices). Whenever a problem fits into this setup, one can massively simplify the calculation by
using an abundance of mathematical structure (see also \tab{tab:ZigZagComplexity}).
The power of parametric integration is its universality. It can be used in many problems
for which graphical functions do not apply.

\begin{table}[t]
\begin{tabular}{|c|c|}
\hline
Program                                      & Timing (CPU core s) \\ \hline
\texttt{HyperInt}           & 900 s  \\ \hline
\texttt{HyperFORM}          & 360 s  \\ \hline
\texttt{HyperlogProcedures} & 10 s   \\ \hline
\texttt{Forcer}             & 0.4 s  \\ \hline
\texttt{Mincer}             & 0.01 s \\ \hline
\end{tabular}
\caption{\label{tbl:FAtimings}Timings (meant for a single CPU core) taken with various programs and approaches for the three-loop p-integral \texttt{FA}.}
\end{table}

A word of caution: Although the user must provide an expression name for the integrand -- and the code internally uses this name, as well as names for all temporary expressions -- it is nevertheless good practice to separate the \form\ sources for each distinct integrand the user wishes to compute. 
 This is because the integration order is fixed by the variable \texttt{IntegrationSequence}, which depends on the particular topology under consideration. As a result, different topologies should not be processed in parallel in the same \form\ source.

%----------------------------------------
\subsection{A non-QFT example}
%----------------------------------------
The \ourcode\ package can be used for any linearly reducible integral. 
To illustrate this, we consider the integral in Eq.~(D.45) of Ref.~\cite{Anastasiou:2013srw}:
\begin{align}
\label{eq:babisD45}
  \mathcal{I}(\ep)
  &=
  \int_{0}^{\infty}
    \ud x_1\int_{0}^{\infty} \ud x_2\int_{0}^{\infty} \ud x_3\,
    x_{1}^{\ep}
    (1 + x_{1})^{-2 + 3\ep}
    x_{2}^{-\ep}
    (1 + x_{2})^{-2 - 4\ep}
    x_{3}^{2 \ep}
    (1 + x_{3})^{-1 - \ep}\,
    \cdot 
    \nonumber\\
    &\qquad\qquad\qquad\qquad\qquad\quad
    \cdot\, (1 + x_{2} + x_{3} + x_{1} x_{3})^{-1 - 2\ep}
  \,.
\end{align}
The integral $I(\ep)$ is \ep-finite and affine; we do not have to use the Chen-Wu theorem.
In order to compute the result published in Ref.~\cite{Anastasiou:2013srw}, the following code can be used:
\begin{verbatim}
#include- hyperform.h

#define HYPMAXEP "3"
#define IntegralExpr "AnastasiouEtAl"
#define IntegrationSequence "1,2,3"

symbols x1,...,x3;
symbol ep;
cfunctions num,den;

local `IntegralExpr' =
  x1^ep *
  (1 + x1)^(3*ep - 2) *
  x2^-ep *
  (1 + x2)^(-4*ep - 2) *
  x3^(2*ep) *
  (1 + x3)^(-ep - 1) *
  (1 + x2 + x3 + x1*x3)^(-2*ep - 1)
;
.sort

#call HypParseInputExpr(ep,num,den,x1,...,x3)
.sort

#call HypAutoRegularize(`IntegralExpr')

#call HypEpExpand

#do IntVar={`IntegrationSequence'}
  #call HypIntegrationStep(`IntegralExpr',`IntVar')
#enddo

#call HypFinalizeResult(ep,`HYPMAXEP')

#call HypPrintStatistics
.sort

bracket ep;
print +s;
.end
\end{verbatim}
The structure of the code is similar to the calculation of FA topology discussed in Sec.~\ref{sec:FAtopo} and 
the order of the calls to the \ourcode procedures is the same.
Here, we expand in \ep\ up to $\mathcal{O}(\ep^3)$.

The result for $I(\ep)$ in  Eq.~(\ref{eq:babisD45}) is then given by
\begin{verbatim}
AnastasiouEtAl =
    + ep * (
       + 5/3
       - 5*z3
       + 4/3*z2
       )
    + ep^2 * (
       - 157/6
       + 745/12*z4
       - 10*z3
       - 32/3*z2
       )
    + ep^3 * (
       + 1175/12
       + 745/6*z4
       - 910/3*z5
       + 607/6*z3
       + 58*z2
       - 277/3*z2*z3
       )
    - 2/3
       + 2/3*z2
      ;
\end{verbatim}
which agrees with the result in Eqs.~(D.47), (D.65) and (D.66) of Ref.~\cite{Anastasiou:2013srw}.

%%%%%%%%%%%%%%%%%%%%%%%%%%%%%%%%%%%%%%%%%%%%%%%%%%%%%
\section{An example at the current limit of computational complexity}
%%%%%%%%%%%%%%%%%%%%%%%%%%%%%%%%%%%%%%%%%%%%%%%%%%%%%
We aim to make full use of \form\!'s performance and present an example that demonstrates its capabilities at the highest level. 
The example derives from current challenges in perturbative quantum field theory, where computations at increasingly higher loops demand both efficiency and scalability. 

%----------------------------------------
\subsection{Zigzag diagrams}
%----------------------------------------
\begin{figure}[ht]
    \centering
    \begin{tikzpicture}
      [vertex/.style={circle,
                      inner sep=0pt,
                      minimum size=1mm,
                      fill}
      ]
      \node (1) [vertex, label = $1$] {};
      \node (3) [below right = of 1, vertex, label = below:$3$] {};
      \node (2) [above right = of 3, vertex, label = $2$] {};
      \node (4) [below right = of 2, vertex, label = below:$4$] {};
      \node (5) [above right = of 4, vertex, label = $5$] {};
      \node (6) [below right = of 5, vertex, label = right:$6$] {};
      \node (7) [above right = of 6, vertex, label = $7$] {};
      \draw[-] (1) -- (2) node[above, pos = 0.5] () {$\alpha_1$};
      \draw[-] (1) -- (3) node[above right, pos = 0.5, xshift = -5pt] () {$\alpha_2$};
      \draw[-] (2) -- (3) node[below right, pos = 0.5, xshift = -5pt] () {$\alpha_4$};
      \draw[-] (2) -- (4) node[above right, pos = 0.5, xshift = -5pt] () {$\alpha_5$};
      \draw[-] (2) -- (5) node[above, pos = 0.5] () {$\alpha_6$};
      \draw[-] (3) -- (4) node[below, pos = 0.5] () {$\alpha_7$};
      \draw[-] (4) -- (5) node[below right, pos = 0.5, xshift = -5pt] () {$\alpha_8$};
      \draw[-] (4) -- (6) node[below, pos = 0.5] () {$\alpha_9$};
      \draw[-] (5) -- (6) node[above right, pos = 0.5, xshift = -5pt] () {$\alpha_{10}$};
      \draw[-] (5) -- (7) node[above, pos = 0.5] () {$\alpha_{11}$};
      \draw[-] (6) -- (7) node[above left, pos = 0.5, xshift = +5pt] () {$\alpha_{12}$};
      \draw[-] (1) edge[bend right = 90, looseness = 1.0] node[below] {$\alpha_3$} (7);
    \end{tikzpicture}
    \caption{The six-loop zigzag diagram with vertices labeled by numbers and edges by 
    Schwinger variables.}
    \label{fig:Zigzag6}
\end{figure}
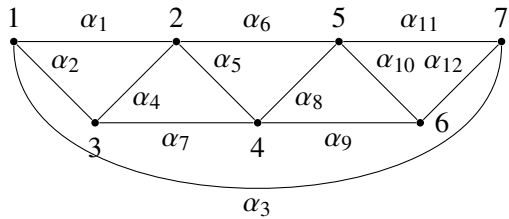

The zigzag family is a set of quite special Feynman diagrams.
Their residues ($p$-integrals) are
\ep-finite and linearly reducible \cite{brown2010periodsfeynmanintegrals}.
Their integrals were conjectured by David Broadhurst and Dirk Kreimer~\cite{Broadhurst:1995km}
and much later proven in Ref.~\cite{Brown:2015ztw} for an arbitrary loop order. 
More recently, an alternative proof was presented in Ref.~\cite{Derkachev:2023bpz}.

In Fig.~\ref{fig:Zigzag6} we depict the six-loop $Z_{6}$ topology with  labels for its vertices and edges.
The Feynman residue for the zigzag diagrams can be written in parametric form 
\begin{align}
\label{eq:zigzag}
    I_{Z_{n}} &=
    \int_{0}^{\infty}\ud x_{1}\cdots \int_{0}^{\infty}\ud x_{2n}\,
    \frac{\delta(x_i-1)
    }{\mathcal{U}_{Z_{n}}(x)^2}
    \,,
\end{align}
where $\mathcal{U}_{Z_{n}}$ is the first Symanzik polynomial. 
The second Symanzik polynomial is absent and therefore the zigzag family has no external momenta.
As projective integrals, one of the integration variables $x_i$ should be fixed to 1, as indicated by the Dirac-delta function.

To assess the complexity of these multiloop diagrams, we depicted the number of Feynman parameters and the number of monomials in the first Symanzik polynomial for the zigzag diagrams until $Z_{10}$
in \tab{tab:ZigZagComplexity} (calculated with \ourufcode \cite{adamkardos_2025_17646663}). 
In the same table we also list the corresponding timings with three different packages to better assess the computational complexity. As pointed out in Sect.\ \ref{pint},
a timing comparison with {\tt HyperlogProcedures} is problematic because it uses a different
mathematical theory.\footnote{
It was pointed out by the referee that the number of monomials in $Z_n$ is
given by $\frac45F_n^2+\frac{n+5}5F_{2n}$, where $F_n$ is the Fibonacci sequence.}

\begin{table}[!th]
    \centering
    \begin{tabular}{|c|c|c|c|c|c|}
      \hline
       Diagram  & No. of variables & No. of monomials  & \texttt{HyperInt} & \ourcode & \texttt{HyperlogProcedures} \\
      \hline
      \hline
        $Z_3$     &  6 &     16 & 0.8 s & 0.05 s & 0.04 s \\
      \hline
        $Z_4$     &  8 &     45 & 1.5 s & 0.3 s & 0.04 s \\
      \hline
        $Z_5$     & 10 &    130 & 54 s & 17 s & 0.04 s \\
      \hline
        $Z_6$     & 12 &    368 & 28 h & 8 h & 0.05 s \\
      \hline
        $Z_7$     & 14 &   1040 & -- & -- & 0.08 s \\
      \hline
        $Z_8$     & 16 &   2919 & -- & -- & 0.13 s \\
      \hline
        $Z_9$     & 18 &   8160 & -- & -- & 0.24 s \\
      \hline
        $Z_{10}$  & 20 &  22715 & -- & -- & 1.2 s\\
      \hline
    \end{tabular}
    \caption{Polynomial complexity of zigzag diagrams in terms of the number of Feynman 
    parameters, the number of monomials in the first Symanzik polynomial and corresponding timings using three different programs.}
    \label{tab:ZigZagComplexity}
\end{table}

Because the zigzag integrals are finite and known but still have abundant polynomial
complexity, they are ideal for testing the performance of \ourcode\!. With the current version of the code we were able to calculate $I_{Z_6}$ and
find a speed-up compared to \hyperint by approximately $20$ times.
The main obstacle for going beyond loop order six lies in the way the code currently
handles polynomials. In the next version of \ourcode\!, this restriction will be lifted.

Zigzag diagrams are linearly reducible for some integration orders but not for
all integration orders. In the next subsection, we illustrate an economic integration
order for the zigzag family. The order is based on an ordering strategy by Brown, that also applies to other Feynman integrals \cite{brown2010periodsfeynmanintegrals}.

%----------------------------------------
\subsection{Good integration sequences}
%----------------------------------------
Good variable sequences for iterated integration are discussed in Ref.~\cite{brown2010periodsfeynmanintegrals}.
The general picture is to complete cycles and vertices: Integration disects a graph $G$
into two parts, the subgraph $G_{\mathrm{done}}$ of the edges that have been integrated
out and the subgraph $G_{\mathrm{todo}}$ of the edges that remain to be integrated.
Isolated vertices are dropped in both graphs. An integration sequence is good if it maximizes
the loops (independent cycles) in $G_{\mathrm{done}}$ and minimizes
the vertices in $G_{\mathrm{todo}}$. In particular, it is valuable to keep
the number of common vertices in $G_{\mathrm{todo}}$ and $G_{\mathrm{done}}$
as small as possible in all steps. In Ref.~\cite{brown2010periodsfeynmanintegrals}
the maximum number of these common vertices is the vertex width of the sequence.
The vertex width of the graph is the minimum vertex width of all sequences.
It has been proven that every graph with vertex width $\leq3$ is linearly reducible.

\begin{figure}
  \begin{subfigure}[b]{0.32\textwidth}
    \centering
    \begin{tikzpicture}
      \node (1) [vertex] {};
      \node (3) [below right = of 1, vertex] {};
      \node (2) [above right = of 3, vertex] {};
      \node (4) [below right = of 2, vertex] {};
      \node (5) [above right = of 4, vertex] {};
      \node (6) [below right = of 5, vertex] {};
      \draw[-] (1) -- (2) node[above, pos = 0.5] {$1$};
      \draw[-] (1) -- (3) node[above, pos = 0.5] {$2$};
      \draw[-] (2) -- (3) node[above, pos = 0.5] {$3$};
      \draw[-] (2) -- (4) node[above, pos = 0.5] {$4$};
      \draw[-] (2) -- (5) node[above, pos = 0.5] {$5$};
      \draw[-] (3) -- (4) node[above, pos = 0.5] {$6$};
      \draw[-] (4) -- (5) node[above, pos = 0.5] {$7$};
      \draw[-] (4) -- (6) node[above, pos = 0.5] {$8$};
      \draw[-] (5) -- (6) node[above, pos = 0.5] {$9$};
      \draw[-] (6) edge[bend left = 80, looseness = 0.8] node[above, pos = 0.5] {$10$} (1);
    \end{tikzpicture}
    \caption{\label{fig:Z5diagram} Original $Z_5$ diagram}
  \end{subfigure}
  \begin{subfigure}[b]{0.32\textwidth}
    \centering
    \begin{tikzpicture}
      \node (1) [vertex] {};
      \node (3) [below right = of 1, vertex] {};
      \node (2) [above right = of 3, vertex] {};
      \node (4) [below right = of 2, vertex] {};
      \node (5) [above right = of 4, vertex] {};
      \node (6) [below right = of 5, vertex] {};
      \draw[-] (1) -- (2) node[above, pos = 0.5] {$1$};
      \draw[-] (1) -- (3) node[above, pos = 0.5] {$2$};
      \draw[-] (2) -- (3) node[above, pos = 0.5] {$3$};
      \draw[-] (2) -- (4) node[above, pos = 0.5] {$4$};
      \draw[-] (2) -- (5) node[above, pos = 0.5] {$5$};
      \draw[-] (3) -- (4) node[above, pos = 0.5] {$6$};
      \draw[-] (4) -- (5) node[above, pos = 0.5] {$7$};
      \draw[-] (4) -- (6) node[above, pos = 0.5] {$8$};
      \draw[-] (5) -- (6) node[above, pos = 0.5] {$9$};
      \draw[ChenWuLine] (6) edge[ChenWuLine, bend left = 80, looseness = 0.8] node[gray, above, pos = 0.5] {$10$} (1);
    \end{tikzpicture}
    \caption{\label{fig:Z5chenWuStep} Chen-Wu applied to edge $10$}
  \end{subfigure}
  \begin{subfigure}[b]{0.32\textwidth}
    \centering
    \begin{tikzpicture}
      \node (1) [BrownVx] {};
      \node (3) [below right = of 1, vertex] {};
      \node (2) [above right = of 3, BrownVx] {};
      \node (4) [below right = of 2, vertex] {};
      \node (5) [above right = of 4, vertex] {};
      \node (6) [below right = of 5, vertex] {};
      \draw[LineUnderInt] (1) -- (2) node[above, pos = 0.5] {$1$};
      \draw[-] (1) -- (3) node[above, pos = 0.5] {$2$};
      \draw[-] (2) -- (3) node[above, pos = 0.5] {$3$};
      \draw[-] (2) -- (4) node[above, pos = 0.5] {$4$};
      \draw[-] (2) -- (5) node[above, pos = 0.5] {$5$};
      \draw[-] (3) -- (4) node[above, pos = 0.5] {$6$};
      \draw[-] (4) -- (5) node[above, pos = 0.5] {$7$};
      \draw[-] (4) -- (6) node[above, pos = 0.5] {$8$};
      \draw[-] (5) -- (6) node[above, pos = 0.5] {$9$};
      \draw[ChenWuLine] (6) edge[ChenWuLine, bend left = 80, looseness = 0.8] node[gray, above, pos = 0.5] {$10$} (1);
    \end{tikzpicture}
    \caption{\label{fig:Z5step1} Edge 1 is integrated out}
  \end{subfigure}
  \begin{subfigure}[b]{0.32\textwidth}
    \centering
    \begin{tikzpicture}
       \node (1) [BrownVx] {};
      \node (3) [below right = of 1, BrownVx] {};
      \node (2) [above right = of 3, BrownVx] {};
      \node (4) [below right = of 2, vertex] {};
      \node (5) [above right = of 4, vertex] {};
      \node (6) [below right = of 5, vertex] {};
      \draw[IntOutLine] (1) -- (2) node[above, pos = 0.5] {$1$};
      \draw[LineUnderIntRight] (1) -- (3) node[above, pos = 0.5] {$2$};
      \draw[-] (2) -- (3) node[above, pos = 0.5] {$3$};
      \draw[-] (2) -- (4) node[above, pos = 0.5] {$4$};
      \draw[-] (2) -- (5) node[above, pos = 0.5] {$5$};
      \draw[-] (3) -- (4) node[above, pos = 0.5] {$6$};
      \draw[-] (4) -- (5) node[above, pos = 0.5] {$7$};
      \draw[-] (4) -- (6) node[above, pos = 0.5] {$8$};
      \draw[-] (5) -- (6) node[above, pos = 0.5] {$9$};
      \draw[ChenWuLine] (6) edge[ChenWuLine, bend left = 80, looseness = 0.8] node[gray, above, pos = 0.5] {$10$} (1);
    \end{tikzpicture}
    \caption{\label{fig:Z5step2} Edge 2 is integrated out}
  \end{subfigure}
 \begin{subfigure}[b]{0.32\textwidth}
    \centering
    \begin{tikzpicture}
%      \node (1) [IntOutVx] {};
       \node (1) [BrownVx] {};
      \node (3) [below right = of 1, BrownVx] {};
      \node (2) [above right = of 3, BrownVx] {};
      \node (4) [below right = of 2, vertex] {};
      \node (5) [above right = of 4, vertex] {};
      \node (6) [below right = of 5, vertex] {};
      \draw[IntOutLine] (1) -- (2) node[above, pos = 0.5] {$1$};
      \draw[IntOutLine] (1) -- (3) node[above, pos = 0.5] {$2$};
      \draw[LineUnderInt] (2) -- (3) node[above, pos = 0.5] {$3$};
      \draw[-] (2) -- (4) node[above, pos = 0.5] {$4$};
      \draw[-] (2) -- (5) node[above, pos = 0.5] {$5$};
      \draw[-] (3) -- (4) node[above, pos = 0.5] {$6$};
      \draw[-] (4) -- (5) node[above, pos = 0.5] {$7$};
      \draw[-] (4) -- (6) node[above, pos = 0.5] {$8$};
      \draw[-] (5) -- (6) node[above, pos = 0.5] {$9$};
      \draw[ChenWuLine] (6) edge[ChenWuLine, bend left = 80, looseness = 0.8] node[gray, above, pos = 0.5] {$10$} (1);
    \end{tikzpicture}
    \caption{\label{fig:Z5step3} Edge 3 is integrated out}
  \end{subfigure}
  \begin{subfigure}[b]{0.32\textwidth}
    \centering
    \begin{tikzpicture}
%      \node (1) [IntOutVx] {};
       \node (1) [BrownVx] {};
      \node (3) [below right = of 1, IntOutVx] {};
      \node (2) [above right = of 3, BrownVx] {};
      \node (4) [below right = of 2, BrownVx] {};
      \node (5) [above right = of 4, vertex] {};
      \node (6) [below right = of 5, vertex] {};
      \draw[IntOutLine] (1) -- (2) node[above, pos = 0.5] {$1$};
      \draw[IntOutLine] (1) -- (3) node[above, pos = 0.5] {$2$};
      \draw[IntOutLine] (2) -- (3) node[above, pos = 0.5] {$3$};
      \draw[-] (2) -- (4) node[above, pos = 0.5] {$4$};
      \draw[-] (2) -- (5) node[above, pos = 0.5] {$5$};
      \draw[LineUnderIntRight] (3) -- (4) node[above, pos = 0.5] {$6$};
      \draw[-] (4) -- (5) node[above, pos = 0.5] {$7$};
      \draw[-] (4) -- (6) node[above, pos = 0.5] {$8$};
      \draw[-] (5) -- (6) node[above, pos = 0.5] {$9$};
      \draw[ChenWuLine] (6) edge[ChenWuLine, bend left = 80, looseness = 0.8] node[gray, above, pos = 0.5] {$10$} (1);
    \end{tikzpicture}
    \caption{\label{fig:Z5step4} Edge 6 is integrated out}
  \end{subfigure}
  \begin{subfigure}[b]{0.32\textwidth}
    \centering
    \begin{tikzpicture}
%      \node (1) [IntOutVx] {};
       \node (1) [BrownVx] {};
      \node (3) [below right = of 1, IntOutVx] {};
      \node (2) [above right = of 3, BrownVx] {};
      \node (4) [below right = of 2, BrownVx] {};
      \node (5) [above right = of 4, vertex] {};
      \node (6) [below right = of 5, vertex] {};
      \draw[IntOutLine] (1) -- (2) node[above, pos = 0.5] {$1$};
      \draw[IntOutLine] (1) -- (3) node[above, pos = 0.5] {$2$};
      \draw[IntOutLine] (2) -- (3) node[above, pos = 0.5] {$3$};
      \draw[LineUnderInt] (2) -- (4) node[above, pos = 0.5] {$4$};
      \draw[-] (2) -- (5) node[above, pos = 0.5] {$5$};
      \draw[IntOutLine] (3) -- (4) node[above, pos = 0.5] {$6$};
      \draw[-] (4) -- (5) node[above, pos = 0.5] {$7$};
      \draw[-] (4) -- (6) node[above, pos = 0.5] {$8$};
      \draw[-] (5) -- (6) node[above, pos = 0.5] {$9$};
      \draw[ChenWuLine] (6) edge[ChenWuLine, bend left = 80, looseness = 0.8] node[gray, above, pos = 0.5] {$10$} (1);
    \end{tikzpicture}
    \caption{\label{fig:Z5step5} Edge 4 is integrated out}
  \end{subfigure}
  \begin{subfigure}[b]{0.32\textwidth}
    \centering
    \begin{tikzpicture}
%      \node (1) [IntOutVx] {};
       \node (1) [BrownVx] {};
      \node (3) [below right = of 1, IntOutVx] {};
      \node (2) [above right = of 3, IntOutVx] {};
      \node (4) [below right = of 2, BrownVx] {};
      \node (5) [above right = of 4, BrownVx] {};
      \node (6) [below right = of 5, vertex] {};
      \draw[IntOutLine] (1) -- (2) node[above, pos = 0.5] {$1$};
      \draw[IntOutLine] (1) -- (3) node[above, pos = 0.5] {$2$};
      \draw[IntOutLine] (2) -- (3) node[above, pos = 0.5] {$3$};
      \draw[IntOutLine] (2) -- (4) node[above, pos = 0.5] {$4$};
      \draw[LineUnderIntRight] (2) -- (5) node[above, pos = 0.5] {$5$};
      \draw[IntOutLine] (3) -- (4) node[above, pos = 0.5] {$6$};
      \draw[-] (4) -- (5) node[above, pos = 0.5] {$7$};
      \draw[-] (4) -- (6) node[above, pos = 0.5] {$8$};
      \draw[-] (5) -- (6) node[above, pos = 0.5] {$9$};
      \draw[ChenWuLine] (6) edge[ChenWuLine, bend left = 80, looseness = 0.8] node[gray, above, pos = 0.5] {$10$} (1);
    \end{tikzpicture}
    \caption{\label{fig:Z5step6} Edge 5 is integrated out}
  \end{subfigure}
  \begin{subfigure}[b]{0.32\textwidth}
    \centering
    \begin{tikzpicture}
%      \node (1) [IntOutVx] {};
       \node (1) [BrownVx] {};
      \node (3) [below right = of 1, IntOutVx] {};
      \node (2) [above right = of 3, IntOutVx] {};
      \node (4) [below right = of 2, BrownVx] {};
      \node (5) [above right = of 4, BrownVx] {};
      \node (6) [below right = of 5, vertex] {};
      \draw[IntOutLine] (1) -- (2) node[above, pos = 0.5] {$1$};
      \draw[IntOutLine] (1) -- (3) node[above, pos = 0.5] {$2$};
      \draw[IntOutLine] (2) -- (3) node[above, pos = 0.5] {$3$};
      \draw[IntOutLine] (2) -- (4) node[above, pos = 0.5] {$4$};
      \draw[IntOutLine] (2) -- (5) node[above, pos = 0.5] {$5$};
      \draw[IntOutLine] (3) -- (4) node[above, pos = 0.5] {$6$};
      \draw[LineUnderInt] (4) -- (5) node[above, pos = 0.5] {$7$};
      \draw[-] (4) -- (6) node[above, pos = 0.5] {$8$};
      \draw[-] (5) -- (6) node[above, pos = 0.5] {$9$};
      \draw[ChenWuLine] (6) edge[ChenWuLine, bend left = 80, looseness = 0.8] node[gray, above, pos = 0.5] {$10$} (1);
    \end{tikzpicture}
    \caption{\label{fig:Z5step7} Edge 7 is integrated out}
  \end{subfigure}
  \begin{subfigure}[b]{0.32\textwidth}
    \centering
    \begin{tikzpicture}
%      \node (1) [IntOutVx] {};
       \node (1) [BrownVx] {};
      \node (3) [below right = of 1, IntOutVx] {};
      \node (2) [above right = of 3, IntOutVx] {};
      \node (4) [below right = of 2, IntOutVx] {};
      \node (5) [above right = of 4, BrownVx] {};
      \node (6) [below right = of 5, BrownVx] {};
      \draw[IntOutLine] (1) -- (2) node[above, pos = 0.5] {$1$};
      \draw[IntOutLine] (1) -- (3) node[above, pos = 0.5] {$2$};
      \draw[IntOutLine] (2) -- (3) node[above, pos = 0.5] {$3$};
      \draw[IntOutLine] (2) -- (4) node[above, pos = 0.5] {$4$};
      \draw[IntOutLine] (2) -- (5) node[above, pos = 0.5] {$5$};
      \draw[IntOutLine] (3) -- (4) node[above, pos = 0.5] {$6$};
      \draw[IntOutLine] (4) -- (5) node[above, pos = 0.5] {$7$};
      \draw[LineUnderIntRight] (4) -- (6) node[above, pos = 0.5] {$8$};
      \draw[-] (5) -- (6) node[above, pos = 0.5] {$9$};
      \draw[ChenWuLine] (6) edge[ChenWuLine, bend left = 80, looseness = 0.8] node[gray, above, pos = 0.5] {$10$} (1);
    \end{tikzpicture}
    \caption{\label{fig:Z5step8} Edge 8 is integrated out}
  \end{subfigure}
  \begin{subfigure}[b]{0.32\textwidth}
    \centering
    \begin{tikzpicture}
%      \node (1) [IntOutVx] {};
       \node (1) [BrownVx] {};
      \node (3) [below right = of 1, IntOutVx] {};
      \node (2) [above right = of 3, IntOutVx] {};
      \node (4) [below right = of 2, IntOutVx] {};
      \node (5) [above right = of 4, BrownVx] {};
      \node (6) [below right = of 5, BrownVx] {};
      \draw[IntOutLine] (1) -- (2) node[above, pos = 0.5] {$1$};
      \draw[IntOutLine] (1) -- (3) node[above, pos = 0.5] {$2$};
      \draw[IntOutLine] (2) -- (3) node[above, pos = 0.5] {$3$};
      \draw[IntOutLine] (2) -- (4) node[above, pos = 0.5] {$4$};
      \draw[IntOutLine] (2) -- (5) node[above, pos = 0.5] {$5$};
      \draw[IntOutLine] (3) -- (4) node[above, pos = 0.5] {$6$};
      \draw[IntOutLine] (4) -- (5) node[above, pos = 0.5] {$7$};
      \draw[IntOutLine] (4) -- (6) node[above, pos = 0.5] {$8$};
      \draw[LineUnderInt] (5) -- (6) node[above, pos = 0.5] {$9$};
      \draw[ChenWuLine] (6) edge[ChenWuLine, bend left = 80, looseness = 0.8] node[gray, above, pos = 0.5] {$10$} (1);
    \end{tikzpicture}
    \caption{\label{fig:Z5step9} Edge 9 is integrated out}
  \end{subfigure}
  \begin{subfigure}[b]{0.32\textwidth}
    \centering
    \begin{tikzpicture}
      \node (1) [IntOutVx] {};
      \node (3) [below right = of 1, IntOutVx] {};
      \node (2) [above right = of 3, IntOutVx] {};
      \node (4) [below right = of 2, IntOutVx] {};
      \node (5) [above right = of 4, IntOutVx] {};
      \node (6) [below right = of 5, IntOutVx] {};
      \draw[IntOutLine] (1) -- (2) node[above, pos = 0.5] {$1$};
      \draw[IntOutLine] (1) -- (3) node[above, pos = 0.5] {$2$};
      \draw[IntOutLine] (2) -- (3) node[above, pos = 0.5] {$3$};
      \draw[IntOutLine] (2) -- (4) node[above, pos = 0.5] {$4$};
      \draw[IntOutLine] (2) -- (5) node[above, pos = 0.5] {$5$};
      \draw[IntOutLine] (3) -- (4) node[above, pos = 0.5] {$6$};
      \draw[IntOutLine] (4) -- (5) node[above, pos = 0.5] {$7$};
      \draw[IntOutLine] (4) -- (6) node[above, pos = 0.5] {$8$};
      \draw[IntOutLine] (5) -- (6) node[above, pos = 0.5] {$9$};
      \draw[ChenWuLine] (6) edge[ChenWuLine, bend left = 80, looseness = 0.8] node[gray, above, pos = 0.5] {$10$} (1);
    \end{tikzpicture}
    \caption{$Z_5$ is fully integrated}
  \end{subfigure}
  \caption{Integration sequence for zigzag 5.}
\end{figure}
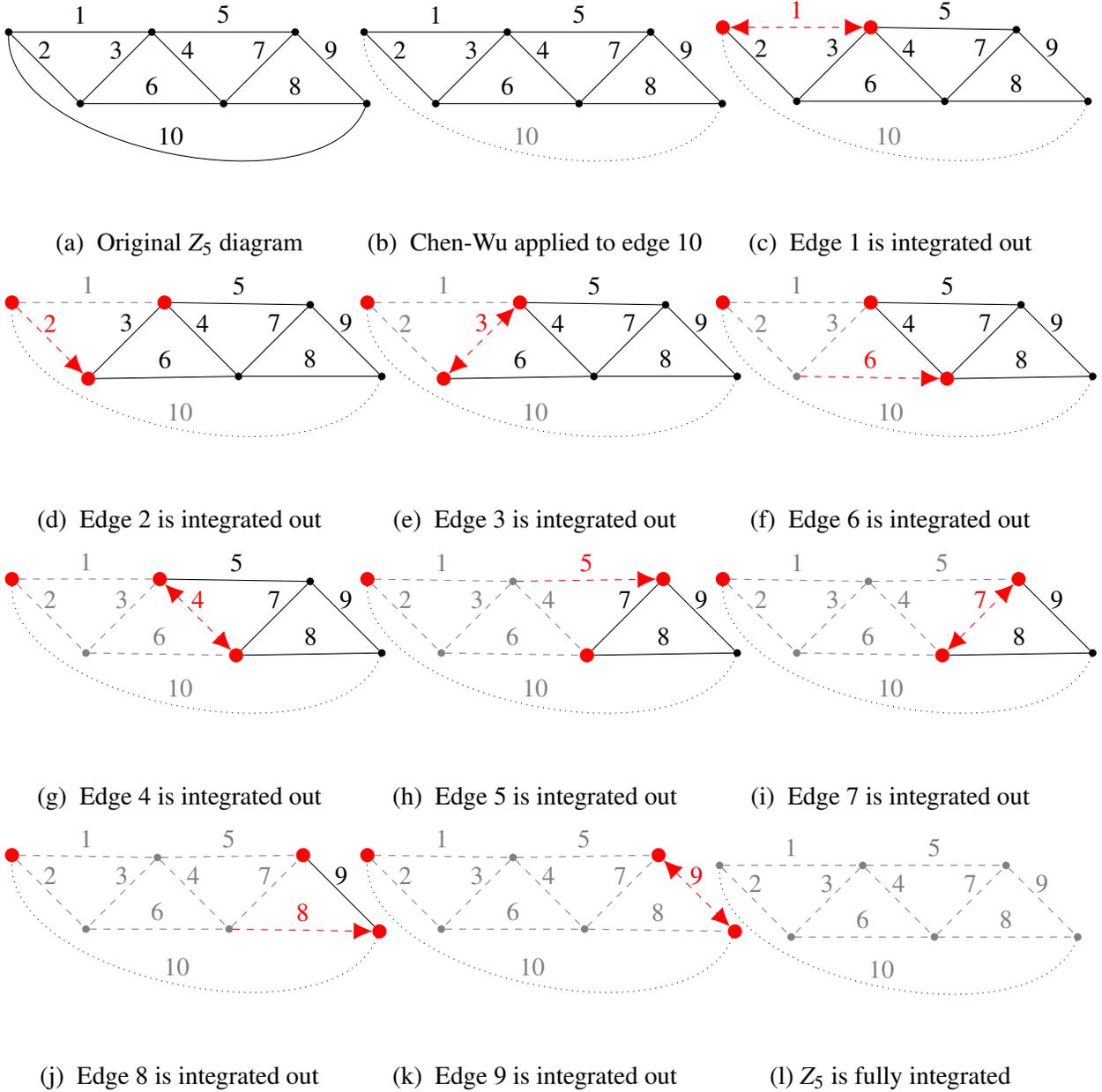

For illustration, we suggest an integration sequence for the zigzag graphs
that has vertex width three.
The case $Z_{5}$ is depicted in Fig.~\ref{fig:Z5diagram}.
Our chosen integration sequence is shown in Figs.~\ref{fig:Z5step1}
-\ref{fig:Z5step9}.
The integration procedure starts with exploiting the Dirac-delta function in Eq.~(\ref{eq:zigzag}), setting one
integration variable to $1$, Fig.~\ref{fig:Z5chenWuStep}. For the 
zigzag family this is always done for the long diagonal.

We integrate out edges $1$, $2$ and $3$ to complete a triangle in $G_{\mathrm{done}}$.
Then, we integrate out edge $6$ to delete a vertex from $G_{\mathrm{todo}}$.
In the next step, we integrate out edge $4$ to complete another triangle in $G_{\mathrm{done}}$.
The procedure continues to mutually delete a vertex in $G_{\mathrm{todo}}$ and to complete
a triangle in $G_{\mathrm{done}}$ until all edges are integrated out.
The graph $Z_5$ has vertex width three as indicated by the red vertices in 
Figs.~\ref{fig:Z5step1}-\ref{fig:Z5step9}.

%----------------------------------------
\subsection{Regularization of divergent integrals}
\label{sec:regularize}
%----------------------------------------
Only very special Feynman integrals are finite in four space-time dimensions. Divergences
generate poles in $\ep$ in dimensional regularization where space-time is generalized
from four to $D = 4 - 2\ep$ dimensions.
A Feynman integral in parametric form cannot be handled with hyperlogarithms if polynomials
are raised to non-integer powers. 
Because the regulator \ep\ is in the exponent, an expansion in \ep\ is mandatory. 
In QFT, there is also no need to know the full \ep\ dependence. 
A naive series expansion in the integrand is only viable for integrals that are \ep-finite,
otherwise a regularization procedure is necessary before expansion. 

The \ep-regularization in Ref.~\cite{Panzer:2014gra} is achieved by an integration-by-parts operation. 
This method is also used in \hyperint.
However, integration-by-parts inherently produces a proliferation in the number and size of terms and is hence inefficient. 
The third author has devised an alternative, more natural regularization procedure that avoids a proliferation of difficulties and that has extensively been used to calculate 'kernel graphical functions'.
The new method will be implemented in a future version of \ourcode; here we stick to the implementation of Ref.~\cite{Panzer:2014gra}.

In any regularization method, one has to detect singularities prior to regularization.
The detection algorithm works as follows.

Consider a parametric integral of the generic form
\begin{align}
  I &=
  \int_{0}^{\infty}\ud x_{1}\cdots \int_{0}^{\infty}\ud x_{N} \integrand(x)
  \,,
\end{align}
where $\mathcal{I}$ is the integrand as a function of Feynman parameters $x=x_{1},\ldots,x_{N}$. 
To determine whether regularization is required, possible singular regions must be identified.  
We do this by considering ordered partitions of a subset of $x$ into variable sets $J$ and $K$.

This is conveniently done with \form using the \texttt{distrib\_} command.

Regularization is generally a multistep procedure, requiring analysis for additional singularities after each step. To identify singular regions, we analyze the scaling behavior of the integrand, where the rescaling is performed 
according to (this is related to the mathematical concept of blowing up singularities)\footnote{There
exist somewhat exotic setups where the rescaling method in this form fails.}
\begin{align}
  x_{j}\in J \implies x_{j}\to \lambda x_{j}
  \,,\quad
  x_{k}\in K \implies x_{k}\to \frac{x_{k}}{\lambda}
  \,,
\end{align}
and the Laurent expansion is carried out in $\lambda$. 
This can be challenging, since after the initial regularization steps, the integrand may become a sum of multiple terms with potential cancellations between them. 
The actual degree of divergence in $\lambda$ may differ from the apparent one due to non-trivial cancellations of leading terms in the Laurent expansion.\footnote{In Feynman integrals, the
singularity structure is given by the coaction of the Connes-Kreimer Hopf algebra of rooted trees \cite{Connes:1999yr}. 
With this tool one can bypass the analysis of singularities by rescaling. \ourcode\!, however,
is designed for wider use, so that we have implemented a general procedure for detecting singularities.}

To properly handle these cancellations, we exploit the fact that for Feynman parameter integrals, the \ep-dependence in the exponents is universal and factorizable. 
Since the regularization procedure involves only manipulations with differential operators, these \ep-dependent exponents remain unaffected throughout the process. 
Thus, even when the integrand consists of multiple terms, we can factorize it according to
\begin{align}
  \integrand(\{x_i\}; \ep) &=
    \mathcal{I}_0(x)
    \,
    \prod_{j=1}^{M} f_j^{m_j \ep}
    =\mathcal{I}_0(x, \ep)\,\mathcal{I}_\ep(x,\ep)\,,
\end{align}
where $\mathcal{I}_0$ changes with each regularization step while $\mathcal{I}_\ep$
is universal. Because $\mathcal{I}_\ep(x,0)=1$, it is insignificant for detecting singular regions.
The residual $\ep$ dependence of $\mathcal{I}_{0}$ is coming from rational function coefficients
depending on $\ep$.

After scaling, we factorize the expression in $\lambda$ which is conveniently achieved
with the \texttt{factarg} command in \form .
The scaling in $\lambda$ is not affected by the other integration variables, masses or momenta. 
Therefore, we choose to substitute them with large primes. 
Then, even in the case of a very complicated $\mathcal{I}_0$, the scaling behavior of $\lambda$ can easily be read off. 
There is a minuscule chance for an accidental cancellation that leads to missing a singular region. Later, in the actual integration steps we check for singularities, so that an error in regularization would be detected.

A region defined by a choice of $J$ and $K$ is divergent if the superficial degree of divergence
\begin{align}
    \omega_{J}^{K}(\integrand) &= |J| - |K| + \deg_{J}^{K}(\mathcal{I}_0)
    \,,
\end{align}
is negative. 
Here, $|J|$ and $|K|$ are cardinalities of $J$ and $K$ (from scaling the integration measure) and $\deg_{J}^{K}(\mathcal{I}_0)$ is the degree of $\mathcal{I}_0$ in $\lambda$ after scaling. 
If a singularity is detected, it will be regularized with integration by parts following Ref.~\cite{Panzer:2014gra}.

If the previous analysis is applied in a recursive way until the set of surviving $J$, $K$ pairs
becomes empty, the original \ep-divergent integrand gets regularized with singularities turned
into \ep\ poles. The resulting integral then becomes suitable for a series expansion in \ep\ allowing integration in the function-space of hyperlogarithms.

%%%%%%%%%%%%%%%%%%%%%%%%%%%%%%%%%%%%%%%%%%%%%%%%%%%%%
\section{Performance metrics}
%%%%%%%%%%%%%%%%%%%%%%%%%%%%%%%%%%%%%%%%%%%%%%%%%%%%%
\begin{figure}
    \centering
    \includegraphics[width=0.8\linewidth]{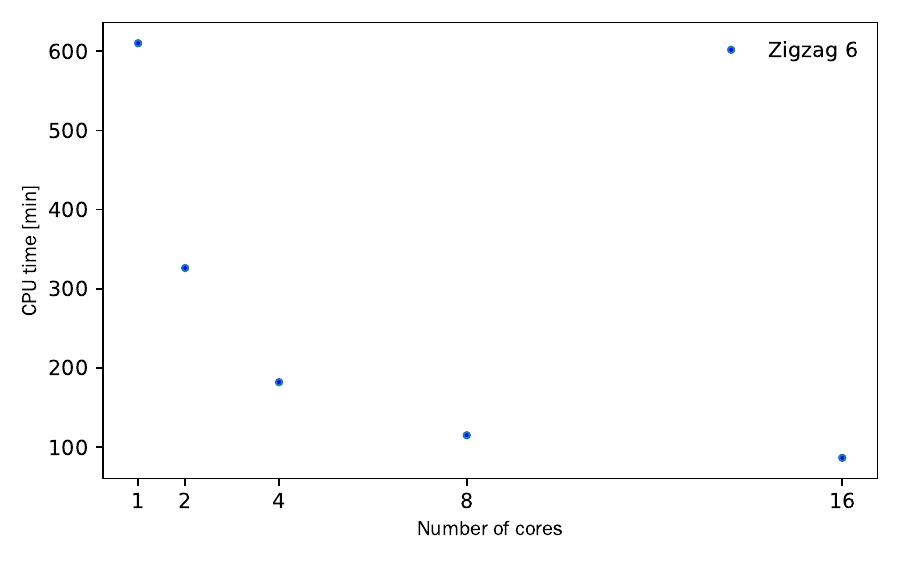}
    \caption{CPU time as a function of the number of \form cores. For details of the computing environment see the main text.}
    \label{fig:number_of_workers}
\end{figure}
\begin{figure}
    \centering
    \includegraphics[width=0.8\linewidth]{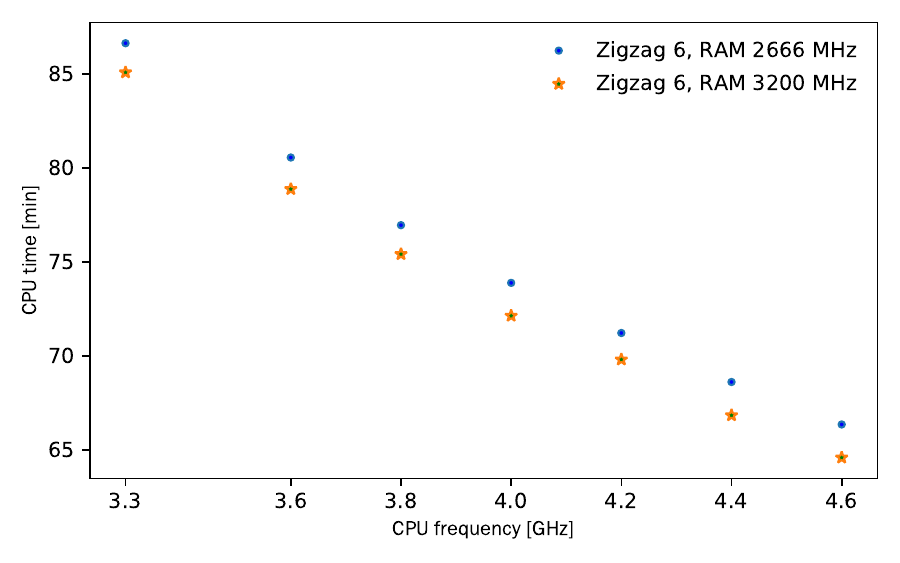}
    \caption{Computation time as a function of CPU clock frequency for two different RAM frequencies, 
    using \form with 16 cores. 
    }
    \label{fig:cpu_frequency}
\end{figure}

\form\! is capable of running on multicore, multi-CPU machines with Unix-like operating systems.
Parallelizing the manipulation of analytic expressions is a highly non-trivial task. Naively we could
assume that running a CAS on $n$ cores would result in a speed-up of $n$. Unfortunately this is almost
always not the case. The speed-up that could be gained on a parallel run strongly depends on the problem
at hand, the nature of manipulations and the structuring of the calculation.

To investigate the scaling of computing time in terms of CPU cores, we ran the zigzag 6 topology on
a machine with 128 GByte of RAM and an AMD Ryzen 9 5900XT CPU running at $3.3$ GHz.
We used the \texttt{time} Linux command to measure the total execution time while varying the number of cores 
between one and 16 (this CPU has 16 physical cores). The runtimes can be read off from \fig{fig:number_of_workers}. 
The CPU time for executing the code does not decrease linearly with the number of cores, and shows a saturation effect. 
Saturation naturally happens, for example, due to distributing and collecting terms or unequal run times in each core for the various modules.

In computationally intensive calculations on computers, we have two parameters on which execution time can mainly depend: the first being the CPU clock frequency with which the CPU fetches, decodes and
executes instructions, while the second is the frequency with which the CPU can read from or write 
to the main memory of the machine. In modern architectures with multiple layers of various cache 
memories, it is highly non-trivial to see how a calculation depends on these parameters. 

The hardware we used for our benchmarks makes it possible to change both the CPU and the 
RAM read/write frequencies (by overclocking). Thus, we also made runs
keeping the number of nodes fixed at 16 and varying the CPU and RAM access frequencies.
The corresponding plot is displayed in \fig{fig:cpu_frequency}. Varying the CPU frequency
leads to a nearly linear speed-up.
Notice, however, that even if we increase the RAM access frequency from $2666$ MHz to $3200$ MHz, 
the speedup is far from the naively expected $20$\%. The fact that the gain in speed is much less
shows that the multilayered cache system provides ample data to keep the cores busy. 
We expect a significantly greater increase in efficiency if the CPU clock frequency is substantially increased.

%%%%%%%%%%%%%%%%%%%%%%%%%%%%%%%%%%%%%%%%%%%%%%%%%%%%%
\section{Conclusions and Outlook}
%%%%%%%%%%%%%%%%%%%%%%%%%%%%%%%%%%%%%%%%%%%%%%%%%%%%%

We have described the \ourcode program, implemented in \form\!, for symbolically integrating hyperlogarithms multiplied by rational functions. It extends previous work in \maple\!, leveraging
\form\!'s efficiency with large expressions. This method is applicable to a variety of problems, 
including the computation of many Feynman integrals. Several examples and relevant benchmarks have 
been presented. Before closing, we briefly discuss limitations of the current implementation and 
outline plans for future extensions.

%----------------------------------------
\subsection{Limitations of the current implementation}
%----------------------------------------
Currently \ourcode\ is restricted to linear reducibility.
If this is not the case, the program quits with an error message claiming the integration is not possible due to the functional form.
In this first version of the code, an error can even occur in linearly reducible integrands because spurious polynomials are not filtered using the Fubini restriction implemented in \hyperint,  see Ref.~\cite{Panzer:2014gra}.
In this case, changing the integration order can sometimes help to perform the integration.

The divergence structure of hyperlogarithms is at most logarithmic. 
This means that, although the final result is well-defined and regular, individual terms encountered during the integration procedure can diverge in the form of $\log(0)$. 
If the integral is well-defined, these divergences should cancel out in the final result.
It is an intrinsic feature of the algorithm that the computation produces terms in intermediate
results that drop out in subsequent calculation steps.
In principle, we could discard singular terms, but to ensure full control over the calculation, the program keeps track of divergent terms. 
If, nevertheless, any such divergences should remain in the final result, an error message is issued to inform the user that the result is unreliable.

In the current version of the code, we use an explicit representation of polynomials. This means that for difficult problems, the terms can become very large. 
If this happens, \form\ terminates with an error message indicating that the maximum term size has been reached.

%----------------------------------------
\subsection{Plans for future extensions}
%----------------------------------------
It is planned to continuously develop \ourcode\!.
The restriction to moderate term sizes will be lifted by restructuring the code. 
It may also be worthwhile to investigate methods to collect the rational coefficients of identical hyperlogarithms, thereby reducing the number of terms and the overall size of the expression.

Another significant improvement will be possible by implementing an efficient method to handle singularities.

The next step is to extend admissible integrands beyond the linearly reducible case. 
Mathematically, it is possible to take the algorithm one step further by allowing the integrand to contain a square root that is quadratic in the next integration variable. 
Such a term is still considered rational in the sense of algebraic geometry, as it is possible to eliminate the root by a rational change of variables. 
However, this method is extremely inefficient. 
Therefore, we plan to implement a direct method (devised by the third author) to handle such square roots.

Finally, another interesting direction for future development could be to exploit the structure of certain integrands in QFT to perform the first integrations analytically before initiating the computer calculation. The effectiveness of this method is illustrated in Refs.~\cite{brown2010periodsfeynmanintegrals,Brown:2010bw}. An analytic calculation can avoid handling large rational functions, especially in dimensions $D \geq 6$, where the Symanzik polynomials exhibit higher powers in the parametric representation.

\appendix
%%
%% ---------------------------------------------------------------------------
%%

%%%%%%%%%%%%%%%%%%%%%%%%%%%%%%%%%%%%%%%%%%%%%%%%%%%%%
\section*{Acknowledgments}
%%%%%%%%%%%%%%%%%%%%%%%%%%%%%%%%%%%%%%%%%%%%%%%%%%%%%
The authors are grateful for many useful and inspiring discussions with Francis Brown and Erik Panzer.
Moreover, the authors are indebted to Jos Vermaseren and all developers of \form\ for creating such
a powerful framework for high-energy theory calculations.

A.K. and S.M. acknowledge the ERC Advanced Grant 101095857 {\it Conformal-EIC}, and 
O.S. has been supported by the DFG through grant SCHN 1240/3-1.
A.K. acknowledges the EIC Theory Institute of the Brookhaven National Laboratory, where this research was partially completed. A.K. is supported by the University of Debrecen Program for Scientific Publication.

\appendix
%%%%%%%%%%%%%%%%%%%%%%%%%%%%%%%%%%%%%%%%%%%%%%%%%%%%%
\section{File and procedure listings}
%%%%%%%%%%%%%%%%%%%%%%%%%%%%%%%%%%%%%%%%%%%%%%%%%%%%%

\subsection{Sources of \ourcode}
In this section, we collect the files in \ourcode\ and list the functionality
of their main procedures.

\ourcode\ consists of three \form\ source files:
\begin{itemize}
    \item \texttt{declare-hyperform.h}
    \item \texttt{hyperform.h}
    \item \texttt{mzvlow.h}
\end{itemize}

\texttt{declare-hyperform.h} contains the declarations of our symbols,
functions, tables and preprocessor variables used throughout the package. All our functions
and symbols have a \texttt{HYP} prefix to ensure separation from user-defined constructs.
Our procedures start with the prefix \texttt{Hyp} to be completely distinguishable from other
user-defined procedures or from other \form\ packages in \texttt{FORMPATH}.
The file begins with the preprocessor variables which govern the operations of the package.
These are restricted for development and debugging purposes.
The user -- under normal circumstances -- does not have to alter their values.

\texttt{hyperform.h} is the main source file of \ourcode. It contains all procedures
of {\tt HyperFORM}. 
For better book-keeping our procedures are organized into categories depending
on intended functionality and are put inside fold environments to facilitate better and faster
navigation through the source code.

\texttt{mzvlow.h} tabulates all MZV reductions up to weight ten, which suffices for calculations up to loop order six.
The results in this file are taken from \form's {\it Multiple Zeta Values Data Mine}\footref{footnote-mzvlow} where MZV reductions up to weight 22 have been published~\cite{Blumlein:2009cf}.

\subsection{Main procedures of \ourcode}
In this subsection, we collect the procedures that the user can use in various phases
of a calculation with \ourcode.

\subsubsection{\texttt{HypParseInputExpr}}
This procedure should be called with the following syntax:
\begin{verbatim}
    #call HypParseInputExpr(
      <user_epsilon_symbol>,
      <user_numerator_function>,
      <user_denominator_function>,
      <list_of_integration_variables>
    )
\end{verbatim}
It converts an integrand defined by the user to the internal representation of \ourcode. 
An example for the integrand in terms of polynomials, numerator and denominator
functions has been presented in Eq.~(\ref{eq:integrand-example}).
The conversion with \texttt{HypParseInputExpr} is applied to all active expressions. 
Also note that non-integration variables must not be specified in the argument list of this procedure.

\subsubsection{\texttt{HypAutoRegularize}}
This procedure allows one to work with divergent parametric integrals by regularizing them. 
As explained in Sec.~\ref{sec:regularize}, it uses the method outlined in Ref.~\cite{Panzer:2014gra}. 
Given a single-term \ep-divergent
integral, the procedure iteratively regularizes the integrand by revealing the corresponding
poles in \ep. To function properly, the user must start with an integrand consisting of a single term only. 
A basic call to the procedure looks like:
\begin{verbatim}
  #call HypAutoRegularize(
    <user_expression_with_integrand>,
    <optional_arguments_of_possible_constants>
  )
\end{verbatim}
The first argument of the call contains the name of the expression that holds the
possibly \ep-divergent integrand. 
As a good practice, users can include this call in all their calculations. 
If no \ep-divergence is found, the original integrand is kept intact. 
To check for any remaining divergences, the code uses a semi-numerical approach. 
For correct functioning, it must be aware of all non-numerical constants besides the integration variables. 
Users should specify these constants as additional arguments when calling the procedure.

As a simple example, consider the integrand for the two-mass one-loop triangle:
\begin{align}
\label{eq:twomassTri}
    f &=
    (z_1 + z_2 + z_3)^{-1+2\ep}
    \left(
      z_1^2 + z_3^2 + 2 z_1 z_3\left( 1 + \frac{S}{m^2}\right)
    \right)^{-1-\ep}
    \,.
\end{align}
In \form\ syntax, this integrand can be given as a local expression \texttt{F} 
\begin{verbatim}
  local F = ( z1 + z2 + z3 )^(-1+2*ep) * 
        ( z1^2 + z3^2 + 2*z1*z3*( 1 + [S/m^2]) )^(-1-ep) ;
\end{verbatim}
where the ratio of kinematical variables $S/m^2$ is designated in \form\ by \verb+[S/m^2]+.
For this integrand the following call has then to be made:
\begin{verbatim}
  #call HypAutoRegularize(F,[S/m^2])
\end{verbatim}
The algorithm described in  Sec.~\ref{sec:regularize} regularizes the integral by extracting the singularity.
After calling \texttt{HypAutoRegularize} the integrand of two-mass one-loop triangle in Eq.~\eqref{eq:twomassTri} then reads:
\begin{align}
\label{eq:twomassTri-reg}
    f &= -\frac{1-2\ep}{2\ep}(z_1+z_3)
    (z_1 + z_2 + z_3)^{-2+2\ep}
    \left(
      z_1^2 + z_3^2 + 2 z_1 z_3\left( 1 + \frac{S}{m^2}\right)
    \right)^{-1-\ep}
    \, ,
\end{align}
which can then be safely expanded in $\ep$.

\subsubsection{\texttt{HypEpExpand}}
Since we work in $D=4-2\ep$ dimensions and since hyperlogarithms can only be formed from functions with integer exponents, a series expansion in \ep\ is required (after regularization, if necessary). 
Because the integrand may have \ep-dependent exponents and coefficients that are rational functions of \ep, we use a simplified implementation of the series expansion in \ep. 
To expand the integrand in terms of the internal \ep\ representation, {\tt HYPep} in \ourcode, the user only needs the following call:
\begin{verbatim}
  #call HypEpExpand
\end{verbatim}
By default, the depth of the \ep-expansion is set to ${\cal O}(\ep^0)$. This can be changed with the preprocessor variable {\tt HYPMAXEP}.

\subsubsection{\texttt{HypApplyChenWu}}
In the parametric form, Feynman integrals are projective integrals, meaning the integration variables are subject to a constraint (usually denoted by a Dirac-delta function) that reflects projectivity.
As explained in Sec.~\ref{sec:impl}, the Dirac-delta function in the integrand gets evaluated and substitutes one integration variable by one. 
This operation is made possible by the Chen-Wu theorem, implemented in \ourcode\ through the procedure:
\begin{verbatim}
  #call HypApplyChenWu(
    <integrand_expression>,
    <preprocessor_variable_or_number>
  )
\end{verbatim}
In the call, the first argument specifies the local expression to which the substitution should be applied, while the second argument can be either a preprocessor variable or a number. This number, or the value of the preprocessor variable, should correspond to the ordinal position of the integration variable that the user wishes to substitute with one.
For example
\begin{verbatim}
  #define ChenWuVar "3"
  #call HypApplyChenWu(F,`ChenWuVar')
\end{verbatim}
or
\begin{verbatim}
  #call HypApplyChenWu(F,3)
\end{verbatim}
will, in both of these cases, substitute the third integration variable ($z_3$ or, internally, \texttt{HYPz3}) by one. 
This variable should coincide with the third variable that the user specifies as an integration variable when calling \texttt{HypParseInputExpr}.

\subsubsection{\texttt{HypIntegrationStep}}

This is the main procedure carrying out the integration in one variable. This procedure can be
called from a code such as
\begin{verbatim}
  #call HypIntegrationStep(
    <integrand_expression>,
    <ordinal_number_of_current_integration_variable>
  )
\end{verbatim}
where the first argument should coincide with the name of the local expression containing
the integrand. 
The second argument must be a number corresponding to the position of the current integration variable in the \texttt{IntegrationSequence}.

\subsubsection{\texttt{HypToGscheme}}

With multi-loop single scale integrals, it is customary to normalize the integral in terms of self-energies raised to the power of the number of loop momentum integrations. 
In the literature this is commonly referred to as the G-scheme \cite{Chetyrkin:1980pr}. 
To adhere to the convention, we provide the function \texttt{HypGscheme} that converts the integral to this normalization. 
To ensure direct comparability with classic results, all master $p$-integrals in the examples (see App.~\ref{app:examples}) are calculated using that scheme.  
To use the procedure, it should be called as:
\begin{verbatim}
  #call HypToGscheme(
    <number_of_loop_momenta>,
    <gamma_function_identifier>,
    <inverse_gamma_function_identifier>,
    <user_defined_epsilon_symbol>,
    <maximum_power_of_epsilon_in_result>
  )
\end{verbatim}

\subsubsection{\texttt{HypFinalizeResult}}

\ourcode uses several internal variables. In order to present the result in a more human-readable format, \texttt{HypFinalizeResult} can be called. 
The functionality of this routine is three-fold: it introduces the shorthand notation for MZVs (including even-integer MZVs), expands coefficient functions that contain rational functions in \ep, and truncates series expansions at a specified order.
The proper calling sequence for this procedure is:
\begin{verbatim}
  #call HypFinalizeResult(
    <user_defined_epsilon_symbol>,
    <maximum_power_of_epsilon_in_result>
  )
\end{verbatim}

\subsubsection{\texttt{HypFibrationBasis}}
\label{subsec:FibrationBasis}

This is a general-purpose convenience function that can be used to convert limits of hyperlogarithms into a fibration basis. 
Whenever \ourcode\ is applied to a multi-scale problem, the final result will still contain hyperlogarithms regularized at infinity, with the scales or parameters that have not been integrated over remaining in the letters. 
In order to eliminate the dependence on parameters in letters, this routine can be applied with the following arguments:
\begin{verbatim}
  #call HypFibrationBasis(
    <name_of_expression_holding_hyperlogs>,
    <regularized_limit_hyperlog_function_name>,
    <hyperlog_function_name_used_after_conversion>,
    <rational_function_identifier>,
    <list_of_parameters>
  )
\end{verbatim}
The first argument specifies the expression containing the hyperlogarithm(s) to be converted. 
The second argument indicates the function head used for the hyperlogarithms with regularized limits. 
The third argument determines the function head to be used for the resulting hyperlogarithms on the fibration basis--which may involve letters containing rational functions. 
The next argument provides the function head for rational functions, while the final arguments list the parameters of the fibration basis.
Note that the program starts to eliminate the dependence in letters starting with the first parameter
from the left-hand-side of the list.

%%%%%%%%%%%%%%%%%%%%%%%%%%%%%%%%%%%%%%%%%%%%%%%%%%%%%
\section{Basic examples}
\label{app:BasicExamples}
%%%%%%%%%%%%%%%%%%%%%%%%%%%%%%%%%%%%%%%%%%%%%%%%%%%%%
In this section, we illustrate the working of the code using very basic examples.

%%%%%%%%%%%%%%%%%%%%%%%%%%%%%%%%%%%%%%%%%%%%%%%%%%%%%
\subsection{Integrating a square}
\label{app:RatFuncExample}
%%%%%%%%%%%%%%%%%%%%%%%%%%%%%%%%%%%%%%%%%%%%%%%%%%%%%

Consider the following integral:
\begin{align}
    \mathcal{I}_{1} &= \int_{0}^{\infty} \frac{\ud x}{(a + b x)^2}
    \,.
\end{align}
This integral is trivial and evaluates to:
\begin{align}
    \mathcal{I}_{1} &= \frac{1}{a b}
    \,.
\end{align}
The same integral could be calculated with the following \texttt{FORM} code snippet which cannot be 
more simple:
\begin{verbatim}
#include- hyperform.h
#define IntegrationSequence ""
symbols a,b;
local F = HYPden(a + b*HYPz1)^2;
.sort
#call HypIntegrationStep(F,1)
#call HypFinalizeResult(HYPep,0)
print;
.end
\end{verbatim}
The working of the code is the following: the \ourcode package is loaded with the \texttt{\#include} 
statement. The define statement is a mandatory, empty variable which holds the integration variables in order of
integration in more complex cases. We declare two generic parameters ($a$ and $b$). Our integrand is
contained by the expression \texttt{F}. In order to minimize variable declarations we directly use
the internal integration variable and denominator function\footnote{In \texttt{FORM} it is preferred to
put denominators into a dedicated function argument.}. A single integration is performed in expression \texttt{F} for \texttt{HYPz1} (hence the $1$ in the argument of the procedure call). In general we
have $\epsilon$-dependent rational function coefficients, to suppress these a call is made to
\texttt{HypFinalizeResult}. Since in this simple case there is no dimensional regularization parameter
in the first argument its internal representation \texttt{HYPep}, is used, while the second argument
stands for the last survining order in \texttt{HYPep} which is in this case zero. The \texttt{print} statement
is responsible for printing out the result to the standard output. When the code is run the 
following output is produced:
\begin{verbatim}
Time =       0.00 sec    Generated terms =          1
               F         Terms in output =          1
                         Bytes used      =         60

   F =
       + HYPden(a)*HYPden(b)
      ;

  0.00 sec out of 0.00 sec
\end{verbatim}

%%%%%%%%%%%%%%%%%%%%%%%%%%%%%%%%%%%%%%%%%%%%%%%%%%%%%
\subsection{Integrating a product}
\label{app:RatFuncExample2}
%%%%%%%%%%%%%%%%%%%%%%%%%%%%%%%%%%%%%%%%%%%%%%%%%%%%%

Consider the following integral:
\begin{align}
    \mathcal{I}_{2} &= \int_{0}^{\infty} \frac{\ud x}{(a + b x)(c + d x)}
    \,.
\end{align}
This integral evaluates to:
\begin{align}
    \mathcal{I}_{2} &= \frac{\log\frac{ad}{bc}}{ad-bc}
    \,.
\end{align}
This integral can also be calculated within \form\ using the \ourcode{} package. A 
minimal code achieving this is listed below:
\begin{verbatim}
#include- hyperform.h
#define IntegrationSequence ""
symbols a,b,c,d;
cfunction log,L,Linf,rat,den;
local F = HYPden(a + b*HYPz1)*HYPden(c+d*HYPz1);
.sort
#call HypIntegrationStep(F,1)
#call HypFinalizeResult(HYPep,0)
multiply replace_(HYPden,den,HYPrat,rat,HYPLinfRegInfZero,Linf);
#call HypFibrationBasis(F,Linf,L,rat,a,b,c,d)
id L(0,a?) = log(a);
bracket den;
print +s;
.end
\end{verbatim}
This code only slightly differs from the previous one. This time we used \form's 
built-in function \texttt{multiply} to perform a multiplication of all the terms with 
the argument, which is also a \form\ built-in function. The arguments of \texttt{replace\_} 
are coming in pairs: the first one gets replaced by the second, etc. With the help
of this line of code the output can be quickly converted from the detailed internal notation
to a more compact, better user-readable format. In order to get the result in terms of
hyperlogarithms where the variable dependence is specific \ourcode{} provides a procedure
to calculate the fibration basis (for a detailed description of the arguments we refer
the reader to Sec.~\ref{subsec:FibrationBasis}). The resulting expression will be given
as a function of hyperlogarithms using our notation of:
\begin{align*}
    L(w; z) &= \mathtt{L}(w, z)
    \,,
\end{align*}
where the left hand side is our mathematical notation while the right hand side shows the role of the different
arguments. After calculating the fibration basis the resulting hyperlogarithms are 
turned into logarithms using an \texttt{id}(\texttt{identity}) statement and Eq.~\eqref{eq:HyperlogToLogConversion}.
The final \texttt{bracket} statement collect terms using their denominators.

When the code is run the following output is produced:
\begin{verbatim}
   F =

       + den( - b*c + a*d) * (
          + log(a)
          - log(b)
          - log(c)
          + log(d)
          );

  0.03 sec out of 0.04 sec
\end{verbatim}

%%%%%%%%%%%%%%%%%%%%%%%%%%%%%%%%%%%%%%%%%%%%%%%%%%%%%
\section{Integrals with multiple scales}
\label{app:multiscale}
%%%%%%%%%%%%%%%%%%%%%%%%%%%%%%%%%%%%%%%%%%%%%%%%%%%%%
The \ourcode{} package also allows the evaluation of integrals that contain many free parameters. In the context of Feynman integrals, this situation arises when multiple physical scales are present. For this reason, we colloquially refer to such cases as integrals with multiple scales. 
For a definite example consider the massless one-loop four-point function:
\begin{align}
    I &=
    \ue^{\ep\EulerG}\,
      \mu^{2\ep}\,
      \int 
      \frac{\ud^D \ell}{i\pi^{D/2}}\, 
      \frac{1}
           {\ell^2 \left(\ell - p_1\right)^2 \left(\ell - p_{12}\right)^2 \left(\ell - p_{123}\right)^2}
    \,,\qquad
    p_{ij\dots k} = p_{i} + p_{j} + \dots + p_{k}
    \,,
\end{align}
where all external momenta are light-like. When using parametric integration this integral
becomes:
\begin{align}
    I &=
    \ue^{\ep\EulerG}\, 
    \mu^{2\ep}\, 
    \Gamma(2 + \ep)\, 
    \int\ud x_1\cdots \ud x_4\, \delta(x_i - 1)\, \mathcal{U}^{2\ep}\,\mathcal{F}^{-2-\ep}
    \,,
\end{align}
with the two Symanzik polynomials 
\begin{align}
    \mathcal{U} &= x_1 + \dots + x_4
    \,,\qquad
    \mathcal{F} = -s x_1 x_3 - t x_2 x_4
    \,.
\end{align}
Because \ourcode prefers positive definite denominators we calculate:
\begin{align}
    I &=
    \ue^{\ep\EulerG}\,
    \frac{1}{t^2}\,
    \left(\frac{-t}{\mu^2}\right)^{-\ep}\,
    \Gamma(2 + \ep)\,
    \int\ud x_1\cdots \ud x_4\,\delta(x_i - 1)\,\mathcal{U}^{2\ep}\,\widetilde{\mathcal{F}}^{-2-\ep}
    \,,\qquad
    \widetilde{\mathcal{F}} =  x_1 x_3 \left[\frac{-s}{-t}\right] + x_2 x_4
    \,.
\end{align}
In the actual calculation we introduce a short-hand for the ratio $-s/-t$ and treat it as a single variable
(\texttt{[-s/-t]}). 
This integral is $\ep$-divergent and therefore requires regularization. Before performing the actual integration, we must call our regularization procedure. In the trailing arguments, we need to list all free parameters that appear in the integrand; in this case, we include \texttt{[-s/-t]}:
\begin{verbatim}
#call HypAutoRegularize(`IntegralExpr',[-s/-t])
\end{verbatim}
After regularization we need to expand in $\ep$:
\begin{verbatim}
#call HypEpExpand
\end{verbatim}
After the expansion and applying the Chen-Wu theorem (in this instance we choose $x_4$) the
integration can be carried out in the remaining integration variables just like in the previous
cases.

After the integration, we would like to express the result in terms of logarithms rather than hyperlogarithms. To achieve this, we introduce a fibration basis. In order to minimize the required manipulations, we recommend converting the hyperlogarithm letters so that they include explicit minus signs. In this example, this is done by introducing \texttt{[s/-t]}, and we then perform the fibration basis operation in these two variables.
\begin{verbatim}
#call HypFibrationBasis(`IntegralExpr',Linf,L,rat,[-s/-t],[s/-t])
\end{verbatim}
As in the previous case we refer the reader to Sec.~\ref{subsec:FibrationBasis} for a detailed
explanation of this procedure. 
In multiscale problems, the proliferation of variables and the increased complexity of the rational function coefficients make it more laborious to obtain a compact final result. Without detailing all intermediate manipulations here, we refer the reader to the example included with the code, which illustrates the steps required to reach the final expression. In the present case, the integrations and manipulations were carried out without including the prefactor
\begin{align}
  \mathcal{N} &=
    \ue^{\ep\EulerG}
    \frac{1}{t^2}
    \left(\frac{-t}{\mu^2}\right)^{-\ep}
    \Gamma(2 + \ep)
  \,,
\end{align}
This factor is only included after the lengthy manipulations to simplify the coefficients.
After including the prefactor and performing an analytical continuation (making all arguments
depending on $-s$ instead of $s$) the result is the classic
one,
\begin{align}
  I &=
    \frac{4}{s t \ep^2}
    -
    \frac{2}{s t \ep}
      \left(
        \log\frac{-s}{\mu^2}
        +
        \log\frac{-t}{\mu^2}
      \right)
    +
    \frac{1}{s t}
      \left(
        \log^2\frac{-s}{\mu^2}
        +
        \log^2\frac{-t}{\mu^2}
        -
        \log^2\frac{-s}{-t^2}
        - 8\zeta_2
      \right)
  + \mathcal{O}(\ep)
  \,,
\end{align}
which perfectly agrees with Eq.~(B.12) of Ref.~\cite{Weinzierl:2022eaz}.

%%%%%%%%%%%%%%%%%%%%%%%%%%%%%%%%%%%%%%%%%%%%%%%%%%%%%
\section{Examples}
\label{app:examples}
%%%%%%%%%%%%%%%%%%%%%%%%%%%%%%%%%%%%%%%%%%%%%%%%%%%%%

\ourcode is distributed with a number of examples.
These include multi-loop $p$-integrals, vacuum graphs of the zigzag family, generic parameter integrals 
and fibration basis calculation. In the following we briefly summarize these examples.

\subsection{Basic examples}
%{
These integrals are provided for the convenience of the user to be able to adapt faster to 
the nomenclature of \form:
\begin{itemize}
    \item \texttt{basic/ratfunc.frm}: this is the code we discussed in detail in Sec.~\ref{app:RatFuncExample}.
    \item \texttt{basic/product.frm}: this is our second basic example covered in detail in
    Sec.~\ref{app:RatFuncExample2}.
\end{itemize}
%}

\subsection{$p$-integrals}
%{

The $p$-integral examples are collected in the folder \texttt{examples/master-integrals}.
Note that in all of these examples we consider massless internal lines and the scale is set
to one. The list of the examples is:
\begin{itemize}
  \item \texttt{one-loop/G.frm}: One-loop self-energy illustrated on \figcite{1}{Larin:1991fz}.
  \item \texttt{two-loop/T1.frm}: Two-loop planar $p$-integral, see \figcite{2}{Larin:1991fz}.
  \item \texttt{two-loop/T2.frm}: Two-loop planar $p$-integral with self-energy insertion, 
  calculated without using that it is a convolution of lower loop integrals, see \figcite{4}{Larin:1991fz}.
  \item \texttt{two-loop/T3.frm}: Two-loop planar $p$-integral being iterated one-loop self-energies, calculated without using that it is a product of lower loop integrals, see \figcite{5}{Larin:1991fz}.
  
  \item \texttt{three-loop/BU.frm}: Three-loop planar $p$-integral, see \figcite{9}{Larin:1991fz}.
  \item \texttt{three-loop/FA.frm}: Three-loop planar $p$-integral, see \figcite{10}{Larin:1991fz}.
  \item \texttt{three-loop/LA.frm}: Three-loop planar $p$-integral, see \figcite{7}{Larin:1991fz}.
  \item \texttt{three-loop/O2.frm}: Three-loop planar $p$-integral, see second graph on \figcite{13}{Larin:1991fz}.
  \item \texttt{three-loop/O3.frm}: Three-loop planar $p$-integral composed of a two-loop and one-loop self-energies, calculated brute-force, see third graph on \figcite{13}{Larin:1991fz}.
  \item \texttt{three-loop/O4.frm}: Three-loop planar $p$-integral containing a two-loop self-energy, calculated brute-force, see third graph on \figcite{6}{Larin:1991fz}.
  \item \texttt{three-loop/Y2.frm}: Three-loop planar $p$-integral containing a two one-loop self-energies, calculated brute-force, see middle graph in first row on \figcite{12}{Larin:1991fz}.
  \item \texttt{three-loop/Y5.frm}: Three-loop planar $p$-integral containing a three iterated one-loop self-energies, calculated brute-force, see last graph in second row on \figcite{12}{Larin:1991fz}.
\end{itemize}
%}

\subsection{Scalar integral with multiple scales}
%{
To illustrate the workflow in case a multiscale integral is to be calculated we included
the massless one-loop box integral as an example in the file: \texttt{/examples/multiscale/massless-box.frm}.
%}

\subsection{Some members of the zigzag family}
%{
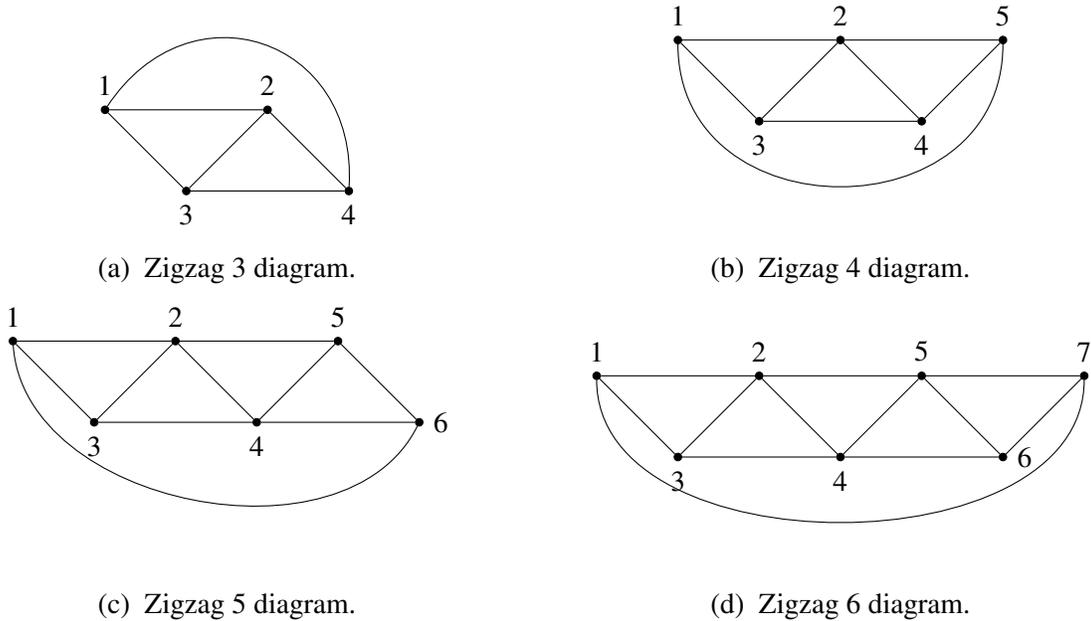
\begin{figure}[ht]
  \begin{subfigure}[t]{0.49\textwidth}
    \centering
    \begin{tikzpicture}
      \node (1) [vertex, label = $1$] {};
      \node (3) [below right = of 1, vertex, label = below:$3$] {};
      \node (2) [above right = of 3, vertex, label = $2$] {};
      \node (4) [below right = of 2, vertex, label = below:$4$] {};
      \draw[-] (1) -- (2);
      \draw[-] (1) -- (3);
      \draw[-] (2) -- (3);
      \draw[-] (2) -- (4);
      \draw[-] (3) -- (4);
      \draw[-] (4) to [bend right = 75, looseness = 1.5] (1);
    \end{tikzpicture}
    \caption{{\label{fig:zigzag3}} Zigzag 3 diagram.}
  \end{subfigure}
  \begin{subfigure}[t]{0.49\textwidth}
    \centering
    \begin{tikzpicture}
      \node (1) [vertex, label = $1$] {};
      \node (3) [below right = of 1, vertex, label = below:$3$] {};
      \node (2) [above right = of 3, vertex, label = $2$] {};
      \node (4) [below right = of 2, vertex, label = below:$4$] {};
      \node (5) [above right = of 4, vertex, label = $5$] {};
      \draw[-] (1) -- (2);
      \draw[-] (1) -- (3);
      \draw[-] (2) -- (3);
      \draw[-] (2) -- (4);
      \draw[-] (2) -- (5);
      \draw[-] (3) -- (4);
      \draw[-] (4) -- (5);
      \draw[-] (5) to [bend left = 90, looseness = 1.5] (1);
    \end{tikzpicture}
    \caption{{\label{fig:zigzag4}} Zigzag 4 diagram.}
  \end{subfigure}
  \begin{subfigure}[t]{0.49\textwidth}
    \centering
    \begin{tikzpicture}
      \node (1) [vertex, label = $1$] {};
      \node (3) [below right = of 1, vertex, label = below:$3$] {};
      \node (2) [above right = of 3, vertex, label = $2$] {};
      \node (4) [below right = of 2, vertex, label = below:$4$] {};
      \node (5) [above right = of 4, vertex, label = $5$] {};
      \node (6) [below right = of 5, vertex, label = right:$6$] {};
      \draw[-] (1) -- (2);
      \draw[-] (1) -- (3);
      \draw[-] (2) -- (3);
      \draw[-] (2) -- (4);
      \draw[-] (2) -- (5);
      \draw[-] (3) -- (4);
      \draw[-] (4) -- (5);
      \draw[-] (4) -- (6);
      \draw[-] (5) -- (6);
      \draw[-] (6) to [bend left = 75, looseness = 1.0] (1);
    \end{tikzpicture}
    \caption{{\label{fig:zigzag5}} Zigzag 5 diagram.}
  \end{subfigure}
  \begin{subfigure}[t]{0.49\textwidth}
    \centering
    \begin{tikzpicture}
      \node (1) [vertex, label = $1$] {};
      \node (3) [below right = of 1, vertex, label = below:$3$] {};
      \node (2) [above right = of 3, vertex, label = $2$] {};
      \node (4) [below right = of 2, vertex, label = below:$4$] {};
      \node (5) [above right = of 4, vertex, label = $5$] {};
      \node (6) [below right = of 5, vertex, label = right:$6$] {};
      \node (7) [above right = of 6, vertex, label = $7$] {};
      \draw[-] (1) -- (2);
      \draw[-] (1) -- (3);
      \draw[-] (2) -- (3);
      \draw[-] (2) -- (4);
      \draw[-] (2) -- (5);
      \draw[-] (3) -- (4);
      \draw[-] (4) -- (5);
      \draw[-] (4) -- (6);
      \draw[-] (5) -- (6);
      \draw[-] (5) -- (7);
      \draw[-] (6) -- (7);
      \draw[-] (7) to [bend left = 90, looseness = 1.0] (1);
    \end{tikzpicture}
    \caption{{\label{fig:zigzag6}} Zigzag 6 diagram.}
  \end{subfigure}
  \caption{The first four zigzag diagrams.}
  \label{fig:Zigzag-diagrams}
\end{figure}

The examples using the zigzag family can be found in the folder \texttt{/examples/zigzag}:
\begin{itemize}
    \item \texttt{zigzag3.frm}: Three-loop zigzag diagram as depicted on \fig{fig:zigzag3}.
    \item \texttt{zigzag4.frm}: Four-loop zigzag diagram as depicted on \fig{fig:zigzag4}.
    \item \texttt{zigzag5.frm}: Five-loop zigzag diagram as depicted on \fig{fig:zigzag5}.
    \item \texttt{zigzag6.frm}: Six-loop zigzag diagram as depicted on \fig{fig:zigzag6}.
\end{itemize}
%}

\subsection{Generic parameter integrals}
%{
These examples are taken from expressions presented in Ref.~\cite{Anastasiou:2013srw}.
They can be found in the folder \texttt{examples/generic}:
\begin{itemize}
    \item \texttt{Anastasiou-Et-Al-eq8p59.frm}: This integral corresponds to Eq.~(8.59) of 
      Ref.~\cite{Anastasiou:2013srw}. 
      In order to directly integrate in terms of hyperlogarithms,
      we have changed the integration limits for all three variables to be between zero and infinity.
    \item \texttt{Anastasiou-Et-Al-eq8p68.frm}: This is given in Eq.~(8.68) of Ref.~\cite{Anastasiou:2013srw}.
      In this case, we have performed a change of integration variables for $x_1$, $x_2$ and $x_3$ in order to have suitable limits for integration in terms of hyperlogarithms.
    \item \texttt{Anastasiou-Et-Al-eq8p69.frm}: This is Eq.~(8.69) of Ref.~\cite{Anastasiou:2013srw}.
      As in the previous case, we also changed the integration variables $x_1$, $x_2$ and $x_3$ to
      better suit our integration procedure.
    \item \texttt{Anastasiou-Et-Al-eqDp45.frm}: This contains Eq.~(D.45) of Ref.~\cite{Anastasiou:2013srw}.
\end{itemize}
%}

\subsection{Fibration basis calculation}
%{
Fibration basis calculation is essential when working with hyperlogarithms. In order to 
demonstrate how this operation could be carried out with \ourcode{} we include an example
in the folder \texttt{examples/fibration-basis}:
\begin{itemize}
    \item \texttt{Anastasiou-Et-Al-eqDp56.frm}: This contains Eq.~(D.56) of 
    Ref.~\cite{Anastasiou:2013srw}. This is the example we have discussed in the main text, 
    \ref{subsec:FibrationBasis}.
\end{itemize}
%}

\smallskip

%\newpage

%

%\bibliographystyle{h-physrev5}

%\begin{thebibliography}{10}
%\end{thebibliography}

\end{document}